\documentclass[fleqn,10pt]{wlscirep}
\usepackage[utf8]{inputenc}
\usepackage[T1]{fontenc}
\usepackage{amssymb,amsthm,amsmath,bbm,bbold,graphicx,epsfig,threeparttable,color}
\usepackage{ulem,enumerate}
\usepackage{enumitem}
\usepackage[title]{appendix}
\usepackage{hyperref}
\hypersetup{colorlinks = true, linkcolor = blue, urlcolor = blue, citecolor = blue}

\theoremstyle{definition}

\title{Effects of symmetry breaking of the structurally-disordered Hamiltonian ensembles on the anisotropic decoherence of qubits}

\author[1,2,*]{Hong-Bin Chen}
\affil[1]{Department of Engineering Science, National Cheng Kung University, Tainan 701401, Taiwan}
\affil[2]{Center for Quantum Frontiers of Research \& Technology, NCKU, Tainan 701401, Taiwan}
\affil[*]{hongbinchen@gs.ncku.edu.tw}

\begin{abstract}
It is commonly known that the dephasing in open quantum systems is due to the establishment of bipartite correlations with ambient environments, which are typically difficult to be fully
characterized. Recently, a new approach of average over disordered Hamiltonian ensemble is developed and shown to be capable of describing the nonclassicality of incoherent dynamics based on inferring
the nonclassical nature of the correlations.
Here we further extend the approach of Hamiltonian ensemble in the canonical form to the realm of structural disorder.
Under the variable separation of the probability distribution within the Hamiltonian ensemble,
the geometrical structure is easily visualized and can be characterized according to the degree of symmetry.
We demonstrate four degrees and investigate the effects of different types of symmetry breaking on the incoherent dynamics.
We show that these effects are easily understood from the emergences of additional terms in the master equations,
leading to rather general master equations and, consequently, going beyond the previous frameworks of pure dephasing or isotropic depolarization.
\end{abstract}
\begin{document}

\maketitle

\section{Introduction}

Exposed to the inevitable interactions with the huge surrounding environments, any quantum systems generically undergo incoherent dynamical processes and gradually lose their quantumness
\cite{breuer_textbook,weiss_textbook,fleming_non_mark_review_annphys_2012,ines_non_mark_review_rmp_2017,hongbin_3sbm_scirep_2015}, constituting the primary obstacle in the developments of frontier
quantum technologies \cite{ladd_quant_comp_nature_2010,buluta_quant_comp_rpp_2011,preskill_quantum_2018,barreiro_q_sim_nature_2011,georgescu_q_sim_rmp_2014,keesling_q_sim_nature_2019,
scully_qhe_pnas_2011,rossnage_1_atom_qhe_science_2016,hongbin_vib_coh_bio_qhe_2016}. Therefore, it is crucial to characterize, control, and eliminate the sources of
decoherence. One of the main causes of the incoherent dynamical nature arises from the damage to the system-environment correlations established during their interactions. However, due to the huge
environmental degree of freedom, it is infeasible to fully access these bipartite correlations. This renders a general, and precise, description of how the correlations are destroyed highly nontrivial.
Consequently, their incoherent effects on the reduced system dynamics are taken into account in terms of a family of completely positive and trace-preserving (CPTP) dynamical linear
maps \cite{kossakowski_cptp_osid_2003,benatti_open_system_dyna_ijmpb_2005,dominy_cptp_maps_qip_2016,chruscinski_open_system_dyna_osid_2017}. Moreover, there are several alternative techniques for
characterizing CPTP maps have been developed, such as the Kraus operators \cite{kraus_textbook}, the process matrices \cite{man_qi_text_book}, and the Choi-Jamio{\l}kowski isomorphism
\cite{jamiolkowski_rmp_1972,choi_laa_1975}.

Besides the system-environment interactions, incoherent dynamics can also arise from a completely different mechanism. Recently, a promising approach, referred to as Hamiltonian ensemble (HE), has been
developed \cite{Kropf2016effective,hongbin_process_n_cla_prl_2018}. HE is initially dedicated to the investigation of disordered systems and classical noises
\cite{Gneiting2016incoherent,kropf2017effective,gneiting_disordered_pra_2017,kropf_disordered_prr_2020}, which are described by an ensemble of Hermitian operators parameterized by some random variables
obeying a specified probability distribution. Irrespective of the unitarity of a single realization generated by each member Hermitian operator, the time-evolved state $\overline{\rho}(t)$
undergoes a dephasing dynamics after the ensemble-averaging procedure over all unitary realizations \cite{Kropf2016effective,hongbin_process_n_cla_prl_2018}. Furthermore, it has been pointed out that
the incoherent dynamical behavior is intimately related to the properties of the probability distribution \cite{Kropf2016effective}. Based on this insight, the probability distribution
encapsulated within the HE has attracted exclusive focus in the characterization of dephasing dynamics, and been promoted to be the canonical Hamiltonian ensemble representation (CHER) of dynamical
processes in the frequency domain \cite{hongbin_process_n_cla_nc_2019,hongbin_cher_sr_2021}. Additionally, the CHER has been shown to be a versatile approach in the quantification of process
nonclassicality \cite{hongbin_process_n_cla_nc_2019}.

However, due to the difficulty imposed by the non-abelian algebraic structure underlying the canonical HE, most of the aforementioned works relied on significant simplifications to circumvent it. For
example, the spectral disorder appealed to HEs consisting of diagonal operators, leading to the pure dephasing \cite{Kropf2016effective}. The process nonclassicality was exemplified
\cite{hongbin_process_n_cla_prl_2018} and quantified \cite{hongbin_process_n_cla_nc_2019} under the same framework of Cartan subalgebra. Furthermore, the attempt going beyond the framework of pure
dephasing was the unitarily invariant disorder, which was studied by incorporating the spectral disorder with the Haar measure integral, leading to the isotropic depolarization
\cite{Kropf2016effective}. Whereas, the most drawback of this unitary invariance approach lies in the deviation from the canonical form, giving rise to the issue of double counting of each member
Hermitian operator. It is also worth noting that there are efforts devoting to the decomposition of pure dephasing into random unitary (RU) representation with static probability distribution
\cite{pernice_randon_unitary_jpb_2012}. However, the generator is still time-dependent, rather than the canonical form.

To cure this issue, as well as to go further beyond the above two frameworks of pure dephasing and isotropic depolarization, here we study the canonical HE of structural disorder along with
different degrees of symmetry. This not only enables us to explore more general types of qubit dephasing dynamics in terms of HE in the canonical form, but also demonstrates the effects of
symmetry breaking of the geometrical structure on the qubit incoherent dynamics. The unitarily invariant disorder is shown to be a special case of the spherical symmetry. We have reduced the
continuous spherical symmetry to three lower levels, until the discrete one of simultaneous reflectional symmetries. Each symmetry breaking gives rise to an additional disturbance complicating the
dynamics, including the anisotropic decay rates, the effective level spacing, and the off-diagonal decay rates. Following this line, the pattern of any further generalization can be deduced.
Finally, we stress that to understand attainable qubit dynamics is important, particulary when we try to find the right type of HE for a physical process.

\section{Hamiltonian ensemble in the spherical coordinates}

We begin with presenting the general HE in the canonical form and exploring the corresponding ensemble-averaged dynamics for a qubit.
Generically, any Hermitian operators acting on a qubit system are elements in the $\mathfrak{u}(2)$ Lie algebra of the form
$\widehat{H}_{\vec{\lambda}}=(\lambda_0\widehat{I}+\lambda_x\hat{\sigma}_x+\lambda_y\hat{\sigma}_y+\lambda_z\hat{\sigma}_z)/2$.
However, as $\lambda_0$ plays no role in describing the qubit dynamics due to the commutativity $[\widehat{I},\hat{\sigma}_j]=0$, we can restrict ourselves to the traceless operators in the
$\mathfrak{su}(2)$ Lie algebra with $\lambda_0=0$.

Here we consider the Hamiltonian ensemble $\{(p_{\vec{\lambda}},\widehat{H}_{\vec{\lambda}})\}_{\vec{\lambda}}$ of canonical form parameterized by $\vec{\lambda}\in\mathbb{R}^3$, where each
member Hamiltonian $\widehat{H}_{\vec{\lambda}}\in\mathfrak{su}(2)$ is associated with a probability $p_{\vec{\lambda}}$ of occurrence. Since the $\mathfrak{su}(2)$ Lie algebra is
non-abelian, the corresponding unitary time-evolution operator $\widehat{U}_{\vec{\lambda}}=\exp(-i\widehat{H}_{\vec{\lambda}}t)$ is rather difficult to be dealt with in these parameters.
Crucially, along with the change of variables to the spherical coordinates $\lambda_x=\omega\sin\theta\cos\phi$, $\lambda_y=\omega\sin\theta\sin\phi$, and $\lambda_z=\omega\cos\theta$, the HE can be
recast into
\begin{equation}
\left\{\left(p(\omega,\theta,\phi),\frac{\omega}{2}\vec{n}\cdot\hat{\sigma}\right)\right\}_{(\omega,\theta,\phi)},
\label{eq_he_sph_coor}
\end{equation}
where $\vec{n}=(\sin\theta\cos\phi,\sin\theta\sin\phi,\cos\theta)\in\mathbb{R}^3$ is the directional unit vector and $\hat{\sigma}$ denotes the three Pauli matrices.
Accordingly, each operator $\widehat{U}_{\vec{\lambda}}=\cos(\omega t/2)\widehat{I}-i\sin(\omega t/2)\vec{n}\cdot\hat{\sigma}$ is explicitly expressed in the spherical coordinates and
leads to a unitarily-evolved single realization
\begin{equation}
\rho_{\vec{\lambda}}(t)=\widehat{U}_{\vec{\lambda}}\rho_0\widehat{U}_{\vec{\lambda}}^\dagger
=\cos^2\frac{\omega t}{2}\rho_0-i\sin\frac{\omega t}{2}\cos\frac{\omega t}{2}\left[\vec{n}\cdot\hat{\sigma},\rho_0\right]+\sin^2\frac{\omega t}{2}(\vec{n}\cdot\hat{\sigma})\rho_0(\vec{n}\cdot\hat{\sigma}),
\label{eq_unitary_single_realization}
\end{equation}
provided an initial state $\rho_0$.

In view of the member Hamiltonian encapsulated within the HE~(\ref{eq_he_sph_coor}), as well as the single realization~(\ref{eq_unitary_single_realization}), we can observe a separation between the
radial coordinate, $\omega$, and the solid angular coordinates, $(\theta,\phi)$. Accordingly, we will further assume a separable probability distribution
\begin{equation}
p(\omega,\theta,\phi)=P(\omega)\Theta(\theta,\phi)
\label{eq_separation_condition}
\end{equation}
into two positive real functions $P(\omega)$ and $\Theta(\theta,\phi)$. We will show that many interesting properties can be conveniently studied with the help of this separation; particularly, the
symmetry is easily exhibited by the geometrical structure of $\Theta(\theta,\phi)$. One should note that, it is $p_{\vec{\lambda}}$ a legitimate probability distribution and normalized to unity
$\int p_{\vec{\lambda}}d^3\vec{\lambda}=\int\int p(\omega,\theta,\phi)\omega^2d\omega d\Omega=1$ with $d\Omega=\sin\theta d\theta d\phi$; while this is not the case for $P(\omega)$ or
$\Theta(\theta,\phi)$ individually. We therefore assume that $\int_0^\infty P(\omega)\omega^2 d\omega=1/\xi$ and $\int\Theta(\theta,\phi)d\Omega=\xi$ due to the separability. This guarantees the normalization condition for $p(\omega,\theta,\phi)$.

Based on the separability, the ensemble-averaged dynamics under the structurally-disordered HE~(\ref{eq_he_sph_coor}) is given by
\begin{eqnarray}
\overline{\rho}(t)&=&\mathcal{E}_t\{\rho_0\}=\int p_{\vec{\lambda}}\widehat{U}_{\vec{\lambda}}\rho_0\widehat{U}_{\vec{\lambda}}^\dagger d^3\vec{\lambda} \nonumber\\
&=&\frac{1}{2}\left(1+\langle\cos\omega t\rangle_P\xi\right)\rho_0
-i\frac{1}{2}\langle\sin\omega t\rangle_P\sum_{j=1}^3\left[\langle n_j\rangle_\Theta\hat{\sigma}_j,\rho_0\right]
+\frac{1}{2}\left(\frac{1}{\xi}-\langle\cos\omega t\rangle_P\right)\sum_{j=1}^3\langle n_j^2 \rangle_\Theta\hat{\sigma}_j\rho_0\hat{\sigma}_j,
\label{eq_averaged_dynamics}
\end{eqnarray}
where, for convenience, we have introduced the radial expectation $\langle f(\omega)\rangle_P=\int_0^\infty f(\omega)P(\omega)\omega^2d\omega$ with respect to $P(\omega)$, $n_j$ the three components of
$\vec{n}$, the 1$^\mathrm{st}$ directional moment $\langle n_j\rangle_\Theta=\int n_j\Theta(\theta,\phi)d\Omega$, and the 2$^\mathrm{nd}$ directional moment
$\langle n_jn_k\rangle_\Theta=\int n_jn_k\Theta(\theta,\phi)d\Omega$. It should be noted that, in the last term of Eq.~(\ref{eq_averaged_dynamics}), there are only three square terms of the
2$^\mathrm{nd}$ directional moment left, due to an appropriate orthogonal transformation. This means that, for any given HE configuration admitting variable separation~(\ref{eq_separation_condition}),
one is possible to redefine a new set of axes of the geometry by using a basis transformation to eliminate the crossing terms $\langle n_j n_k \rangle_\Theta$ with $j\neq k$. Consequently, without loss
of generality, we can start from the dynamical linear map~(\ref{eq_averaged_dynamics}) and ignore the crossing terms in the following. This significantly simplifies the complexity of the problem.
Further details on the orthogonal transformation are discussed in Methods.

On the other hand, the density matrix for a qubit system $\rho=(\widehat{I}+\vec{\rho}\cdot\hat{\sigma})/2$ is also an element in the $\mathfrak{u}(2)$ Lie algebra parameterized by the corresponding
Bloch vector $\vec{\rho}\in\mathbb{R}^3$. Therefore, the properties of the dynamical linear map~(\ref{eq_averaged_dynamics}) can be fully understood from its action on the generators of the
$\mathfrak{u}(2)$ Lie algebra:
\begin{equation}
\left\{\begin{array}{l}
\mathcal{E}_t\{\widehat{I}\}=\widehat{I} \\
\mathcal{E}_t\{\hat{\sigma}_j\}=f_j(t)\hat{\sigma}_j
+\sum_{k,l=1}^{3}\varepsilon_{jkl}\langle\sin\omega t\rangle_P\langle n_l\rangle_\Theta\hat{\sigma}_k, j=1,2,3
\end{array}\right.,
\label{eq_dyn_map_pauli_mat}
\end{equation}
where
\begin{equation}
f_j(t)=\langle\cos\omega t\rangle_P\left(\xi-\langle n_j^2 \rangle_\Theta\right)+\langle n_j^2 \rangle_\Theta/\xi,
\label{eq_f_functions}
\end{equation}
and $\varepsilon_{jkl}=1$ if $(j,k,l)$ is an even permutation of $(1,2,3)$, $-1$ if an odd permutation, and 0 otherwise. From the first line of Eqs.~(\ref{eq_dyn_map_pauli_mat}), we can see that
$\mathcal{E}_t$ is unital. This can be understood if we note that the HE is a special case with time-independent probability distribution and member Hamiltonian of a superset of RU. Due to the
unitality, the decohering behavior of $\mathcal{E}_t$ is of the type of depolarization, captured by the second line of Eqs.~(\ref{eq_dyn_map_pauli_mat}). From the later, we can see that the decohering
behavior of $\mathcal{E}_t$ depends highly on the structure and the symmetry of $p(\omega,\theta,\phi)$ via the radial expectations and the directional moments. On the other hand, this also reflects the
notion of CHER \cite{hongbin_process_n_cla_nc_2019,hongbin_cher_sr_2021}, which conceives the (quasi-)distribution function as the characteristic representation of a dynamics.

Accordingly, we will investigate the dynamical behavior of $\mathcal{E}_t$ along with different degrees of symmetry of the probability distribution exhibited by the solid angular part
$\Theta(\theta,\phi)$.

\section{Spherical symmetry}

\begin{figure*}[htb]
\includegraphics[width=\textwidth]{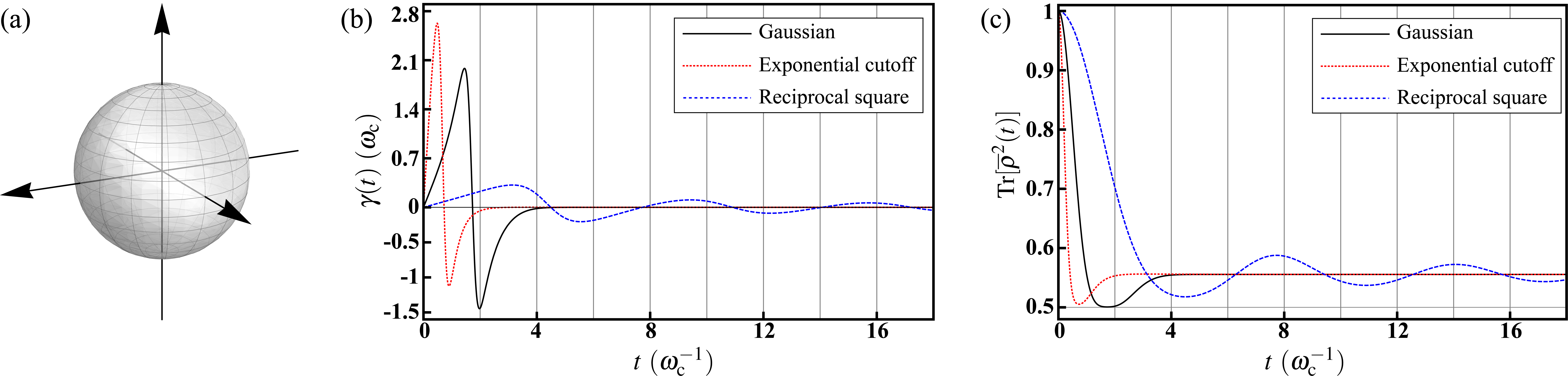}
\caption{Decoherence under the HE with spherically symmetric probability distribution. (a) Visualization of the solid angular part $\Theta(\theta,\phi)=1/4\pi$ of the probability
distribution. As $\Theta(\theta,\phi)$ is a constant, the geometry is a sphere of radius $1/4\pi$. (b) Time-evolution of the decay rates $\gamma(t)$ for the Gaussian (black solid curve), the exponential
cutoff (red dotted curve), and the reciprocal square (blue dashed cruve) radial functions. Their behaviors are different to each other, as explained in the main text. The decay rates temporarily go down
to negative values in some time periods, indicating transitions between indivisibility and full divisibility of the type of random unitary. (c) The purities $\mathrm{Tr}[\overline{\rho}^2(t)]$ governed
by the isotropic depolarization with pure initial states for the three radial functions. There is a rivival in the time period of negative decay rate. Additionally, the final value of the purities is
$5/9$, rather than $1/2$, reflecting that the final state is not the completely mixed state.}
\label{fig_restults_spherical_symm}
\end{figure*}

We first consider the case of spherically symmetric probability distribution with $\Theta(\theta,\phi)=1/4\pi$. Figure~\ref{fig_restults_spherical_symm}a shows its visualization with the distance
between the surface and the origin indicating the value of $\Theta(\theta,\phi)$ in solid angular coordinates $(\theta,\phi)$. As $\Theta(\theta,\phi)$ is a constant, the geometry forms a sphere;
therefore, the HE is of spherical symmetry, which is the highest symmetry can be exhibited. It is straightforward to verify that
$\int_0^\infty P(\omega)\omega^2 d\omega=\int\Theta(\theta,\phi)d\Omega=1$, the 1$^\mathrm{st}$ directional moments $\langle n_j\rangle_\Theta=0$, and the 2$^\mathrm{nd}$ directional moments
$\langle n_j n_k\rangle_\Theta=\delta_{jk}/3$, satisfying the diagonal condition of Eq.~(\ref{eq_averaged_dynamics}).

Under such highly symmetric geometry, from Eqs.~(\ref{eq_dyn_map_pauli_mat}), the ensemble-averaged dynamics is given by
\begin{equation}
\overline{\rho}(t)=\mathcal{E}_t\{\rho_0\}=\left[1-w(t)\right]\frac{\widehat{I}}{2}+w(t)\rho_0,
\end{equation}
which is a statistical mixture between the completely mixed state and the initial state $\rho_0$ along with a time-varying wight $w(t)=(2\langle\cos\omega t\rangle_P+1)/3$. Therefore, the initial state
will gradually lose its coherence. Moreover, this incoherent dynamical behavior is governed by the master equation of isotropic depolarization
\begin{equation}
\frac{\partial}{\partial t}\overline{\rho}(t)=\frac{\gamma(t)}{2}\sum_{j=1}^3\left[\hat{\sigma}_j\overline{\rho}(t)\hat{\sigma}_j-\overline{\rho}(t)\right],
\label{eq_iso_master_eq}
\end{equation}
with decay rate $\gamma(t)=-\dot{w}(t)/2w(t)$; namely, the three Pauli decay channels share identical decay rate. The derivation of the master equation from Eqs.~(\ref{eq_dyn_map_pauli_mat}) is outlined
in Ref.~\cite{Kropf2016effective,lindblad_jmp_1976,andersson_kraus_decompo_jmo_2007,hongbin_k_divi_pra_2015}. Additionally, the decohering dynamics of this isotropic depolarization can also be
well-understood by the purity
\begin{equation}
\mathrm{Tr}[\overline{\rho}^2(t)]=w^2(t)\left[\mathrm{Tr}\left(\rho_0^2\right)-\frac{1}{2}\right]+\frac{1}{2}.
\end{equation}

It is worthwhile to note that, the spherical symmetry qualitatively reproduces the unitarily invariant disorder for qubit~\cite{Kropf2016effective}, which is studied by means of incorporating
the spectral disorder with the Haar measure distributed uniformly over whole solid angular coordinates. However, one of the drawbacks of this approach is the occurrence of double counting of each
member Hamiltonian operator. For example, both of the two identical members $\omega\hat{\sigma}_z/2$ and $\exp(-i\pi\hat{\sigma}_x/2)(-\omega\hat{\sigma}_z/2)\exp(i\pi\hat{\sigma}_x/2)$ contribute
individually to the ensemble-average procedure. On the other hand, the HE in the canonical form rules out this circumstance by considering positive radial coordinate exclusively. Therefore,
quantitative differences can be seen when we further take the radial part $P(\omega)$ into account in the following. To further clarify this situation and to exemplify the isotropic depolarization,
we will carry out several types of distribution functions. As the solid angular part $\Theta(\theta,\phi)$ has been specified according to the symmetry, we will characterize the distribution functions
with the radial part $P(\omega)$.

As a comparative study, we consider the Gaussian radial function defined as
\begin{equation}
P_\mathrm{G}(\omega)= \frac{1}{\omega_\mathrm{c}^3}\sqrt{\frac{2}{\pi}}e^{-\frac{\omega^2}{2\omega_\mathrm{c}^2}}
\label{eq_radial_gaussian}
\end{equation}
for $\omega\geq0$. Note that the standard deviation $\omega_\mathrm{c}$ controls the width of the distribution and therefore plays similar role of cutoff frequency.
The functional form is specified according to the
normalization condition $\int_0^\infty P_\mathrm{G}(\omega)\omega^2d\omega=1$ and therefore the coefficient is slightly different from the usual one. For the Gaussian radial function
$P_\mathrm{G}(\omega)$, one can analytically evaluate
\begin{equation}
\langle\cos\omega t\rangle_{P_\mathrm{G}}=\int_0^\infty \cos\omega t P_\mathrm{G}(\omega)\omega^2d\omega
=e^{-\frac{(\omega_\mathrm{c}t)^2}{2}}\left[1-(\omega_\mathrm{c}t)^2\right].
\label{eq_expectation_cos_gau}
\end{equation}
This allows us to evaluate the mixing weight $w_\mathrm{G}(t)=(2\langle\cos\omega t\rangle_{P_\mathrm{G}}+1)/3$ and the decay rate in Eq.~(\ref{eq_iso_master_eq}):
\begin{equation}
\gamma_\mathrm{G}(t)=\omega_\mathrm{c}\frac{\omega_\mathrm{c}t\left[3-(\omega_\mathrm{c}t)^2\right]}{2\left[1-(\omega_\mathrm{c}t)^2\right]+e^{\left[(\omega_\mathrm{c}t)^2/2\right]}}.
\label{eq_decay_rate_sph_sym_gau}
\end{equation}
If compared with the one obtained by the unitarily invariant disorder for qubit~\cite{Kropf2016effective}, we can find that Eq.~(\ref{eq_decay_rate_sph_sym_gau}) is lesser by a factor 2, indicating
that the HE in the canonical form rules out the circumstance of double counting. The numerical results are shown in Fig.~\ref{fig_restults_spherical_symm}b. It can be seen that, $\gamma_\mathrm{G}(t)$
is initially increasing, then followed by a sharp descent to negative values, and finally approaching zero asymptotically from below. Additionally, the time evolution of the purity with pure initial
state $\mathrm{Tr}[\overline{\rho}^2(t)]=\left[w_\mathrm{G}^2(t)+1\right]/2$ is shown in Fig.~\ref{fig_restults_spherical_symm}c. The purity initially decays very approaching the completely mixed state
due to the increasing $\gamma_\mathrm{G}(t)$; afterwards, it rises again and saturates to the value of $(w_\mathrm{G}^2(t\rightarrow\infty)+1)/2$, indicating that the final state deviates from the
completely mixed state. This property can also be observed in the following two examples and will be discussed latter.


Next, we consider an alternative cutoff in exponential form defined as
\begin{equation}
P_\mathrm{EC}(\omega)=\frac{1}{6\omega_\mathrm{c}^4}\omega e^{-\frac{\omega}{\omega_\mathrm{c}}},
\label{eq_radial_exp_cut}
\end{equation}
satisfying the normalization condition $\int_0^\infty P_\mathrm{EC}(\omega)\omega^2d\omega=1$. $\omega_\mathrm{c}$ is the cutoff value. With this exponential cutoff, we have
\begin{equation}
\langle\cos\omega t\rangle_{P_\mathrm{EC}}=\int_0^\infty \cos\omega t P_\mathrm{EC}(\omega)\omega^2d\omega
=\frac{1-6(\omega_\mathrm{c}t)^2+(\omega_\mathrm{c}t)^4}{\left[1+(\omega_\mathrm{c}t)^2\right]^4},
\label{eq_expectation_cos_exp_cut}
\end{equation}
the mixing weight $w_\mathrm{EC}(t)=(2\langle\cos\omega t\rangle_{P_\mathrm{EC}}+1)/3$, and the decay rate in Eq.~(\ref{eq_iso_master_eq}):
\begin{equation}
\gamma_\mathrm{EC}(t)=4\omega_\mathrm{c}^2t\frac{\left[3-(\omega_\mathrm{c}t)^2\right]\left[1+(\omega_\mathrm{c}t)^2\right]+2\left[1-6(\omega_\mathrm{c}t)^2+(\omega_\mathrm{c}t)^4\right]}
{2\left[1-6(\omega_\mathrm{c}t)^2+(\omega_\mathrm{c}t)^4\right]\left[1+(\omega_\mathrm{c}t)^2\right]+\left[1+(\omega_\mathrm{c}t)^2\right]^5}.
\label{eq_decay_rate_sph_sym_exp_cut}
\end{equation}
From Fig.~\ref{fig_restults_spherical_symm}b, $\gamma_\mathrm{EC}(t)$ exhibits a similar behavior to $\gamma_\mathrm{G}(t)$, but even sharper oscillation. Consequently, this is also the case for the
purity $\mathrm{Tr}[\overline{\rho}^2(t)]=\left[w_\mathrm{EC}^2(t)+1\right]/2$ shown in Fig.~\ref{fig_restults_spherical_symm}c. This is due to the fact that $P_\mathrm{EC}(\omega)$ possesses a long
wing over high $\omega$ domain, leading to a rapid spreading of random unitary rotation and shorter coherence time.


As a third example, we consider a radial function of different type. The reciprocal square
\begin{equation}
P_\mathrm{RS}(\omega)=\frac{1}{\omega_\mathrm{c}\omega^2},
\label{eq_radial_reci_squa}
\end{equation}
for $\omega\in[0,\omega_\mathrm{c}]$, is defined on a finite interval rather than infinite length. Again the functional form is specified by the normalization condition
$\int_0^{\omega_\mathrm{c}}P_\mathrm{RS}(\omega)\omega^2d\omega=1$. Then the radial expectation
\begin{equation}
\langle\cos\omega t\rangle_{P_\mathrm{RS}}=\int_0^{\omega_\mathrm{c}} \cos\omega t P_\mathrm{RS}(\omega)\omega^2d\omega=\frac{\sin\omega_\mathrm{c}t}{\omega_\mathrm{c}t}
\label{eq_expectation_cos_reci_squa}
\end{equation}
leads to the mixing weight $w_\mathrm{RS}(t)=(2\langle\cos\omega t\rangle_{P_\mathrm{RS}}+1)/3$ and the decay rate in Eq.~(\ref{eq_iso_master_eq}):
\begin{equation}
\gamma_\mathrm{RS}(t)=\omega_\mathrm{c}\frac{\sin\omega_\mathrm{c}t-(\omega_\mathrm{c}t)\cos\omega_\mathrm{c}t}{2(\omega_\mathrm{c}t)\sin\omega_\mathrm{c}t+(\omega_\mathrm{c}t)^2}.
\end{equation}
Due to the finite domain of $P_\mathrm{RS}(\omega)$, the decay rate $\gamma_\mathrm{RS}(t)$ exhibits a long oscillating tail as shown in Fig.~\ref{fig_restults_spherical_symm}b. This behavior is very
different from $\gamma_\mathrm{G}(t)$ and $\gamma_\mathrm{EC}(t)$, and renders the purity $\mathrm{Tr}[\overline{\rho}^2(t)]=\left[w_\mathrm{RS}^2(t)+1\right]/2$ oscillating as well after the
initial descent.


It is evident from the above three examples that, whenever the decay rates $\gamma(t)$ go down to negative values, there is a revival of the purity $\mathrm{Tr}[\overline{\rho}^2(t)]$ in the time
period of negative decay rate. We can even clearly observe this phenomenon from the long oscillating tails in the reciprocal square example. This can be explained as a typical transition between
indivisibility and full divisibility \cite{hongbin_k_divi_pra_2015,chruscinski_k_divi_prl_2014,bae_k_divi_prl_2016,hongbin_n_mark_pra_2017} of the type of RU
\cite{chruscinski_k_divi_pra_2015,megier_randon_unitary_sr_2017}.

On the other hand, although the dynamics is governed by the master equation (\ref{eq_iso_master_eq}) of isotropic depolarization, the final state is generically not the completely mixed state. This can
be realized by observing that the purity saturates to a value of $5/9$, rather than $1/2$ in Fig.~\ref{fig_restults_spherical_symm}c. This does not imply the violation of unitality of the
ensemble-averaged dynamics under canonical HE. Since $\langle\cos\omega t\rangle_{\omega^3/3}\rightarrow0$ when $t\rightarrow\infty$ in these examples, we have the steady mixing weight
$w(t\rightarrow\infty)=1/3$ and, consequently, the final state $\overline{\rho}(\infty)=\left[2(\widehat{I}/2)+\rho_0\right]/3$, a constant mixture between the completely mixed state and the initial
state $\rho_0$.

We have shown that the spherically symmetric probability distribution qualitatively reproduces the unitarily invariant disorder for qubit and leads to the master equation of isotropic depolarization
with identical decay rate of the three Pauli decay channels. To demonstrate the versatility of the canonical HE in characterizing the incoherent dynamics beyond isotropic depolarization,
we will further reduce the degree of symmetry and explore its effects on the incoherent dynamical behaviors.

\section{Simultaneous azimuthal and reflectional symmetries}

To release the spherical symmetry, we consider the HE exhibiting simultaneously the azimuthal symmetry and the reflectional symmetry about the $x$-$y$ plane.
Specifically, we will consider the two examples of unbalanced regimes; namely the bagel-shaped $\Theta(\theta,\phi)=\pi^{-2}\sin\theta$ and the dumbbell-shaped $\Theta(\theta,\phi)=(3/4\pi)\cos^2\theta$. Under these symmetries, it is
straightforward to see that the 1$^\mathrm{st}$ directional moments vanish again, $\langle n_j\rangle_\Theta=0$, and the 2$^\mathrm{nd}$ $x$- and $y$-directional moments are equal,
$\langle n_x^2\rangle_\Theta=\langle n_y^2\rangle_\Theta$. However, in contrast to the case of spherical symmetry, they may not necessarily equal to the 2$^\mathrm{nd}$ $z$-directional moment
$\langle n_z^2\rangle_\Theta$; meanwhile, without loss of generality, we assume that the diagonal condition of Eq.~(\ref{eq_averaged_dynamics}) is still hold, i.e., $\langle n_j n_k\rangle_\Theta=0$
for $j\neq k$.

After determining the directional moments according to the symmetries, the action of ensemble-averaged dynamical linear map in Eqs.~(\ref{eq_dyn_map_pauli_mat}) is significantly simplified as
\begin{equation}
\left\{\begin{array}{l}
\mathcal{E}_t\{\widehat{I}\}=\widehat{I} \\
\mathcal{E}_t\{\hat{\sigma}_j\}=f_j(t)\hat{\sigma}_j
, j=1,2,3
\end{array}\right.,
\label{eq_dyn_map_pauli_mat_azi_ref_symm}
\end{equation}
Since the spherical symmetry is no longer hold, the actions of
$\mathcal{E}_t$ on the three generators are different. Consequently, the incoherent dynamical behavior is governed by the master equation of anisotropic depolarization
\begin{equation}
\frac{\partial}{\partial t}\overline{\rho}(t)=\sum_{j=1}^3\frac{\gamma_j(t)}{2}\left[\hat{\sigma}_j\overline{\rho}(t)\hat{\sigma}_j-\overline{\rho}(t)\right],
\label{eq_aniso_master_eq}
\end{equation}
with decay rates $\gamma_j(t)=\left[\dot{f}_j(t)/2f_j(t)\right]-\sum_{k\neq j}\dot{f}_k(t)/2f_k(t)$ associated to the corresponding Pauli decay channels. Moreover, due to the azimuthal symmetry, we have
$\langle n_x^2\rangle_\Theta=\langle n_y^2\rangle_\Theta$, $f_x(t)=f_y(t)$, and, consequently, $\gamma_x(t)=\gamma_y(t)=-\dot{f}_z(t)/2f_z(t)$. Each decay rate can be considered as a competition between
$\left[\dot{f}_j(t)/2f_j(t)\right]$'s, which are determined by the 2$^\mathrm{nd}$ directional moments. Accordingly, we will consider the two unbalanced regimes,
$\langle n_x^2\rangle_\Theta=\langle n_y^2\rangle_\Theta>\xi/3>\langle n_z^2\rangle_\Theta$ and $\langle n_x^2\rangle_\Theta=\langle n_y^2\rangle_\Theta<\xi/3<\langle n_z^2\rangle_\Theta$.
Moreover, the profile of purity now depends on the initial state $\rho_0=(\widehat{I}+\vec{\rho}_0\cdot\hat{\sigma})/2$:
\begin{equation}
\mathrm{Tr}[\overline{\rho}^2(t)]=\frac{1}{2}\left[1+|\vec{\rho}(t)|^2\right]=\frac{1}{2}\left[1+\sum_{j=1}^3|\rho_{0,j}f_j(t)|^2\right].
\end{equation}

\subsection{Bagel-shaped solid angular function}
\begin{figure*}[ht]
\includegraphics[width=\textwidth]{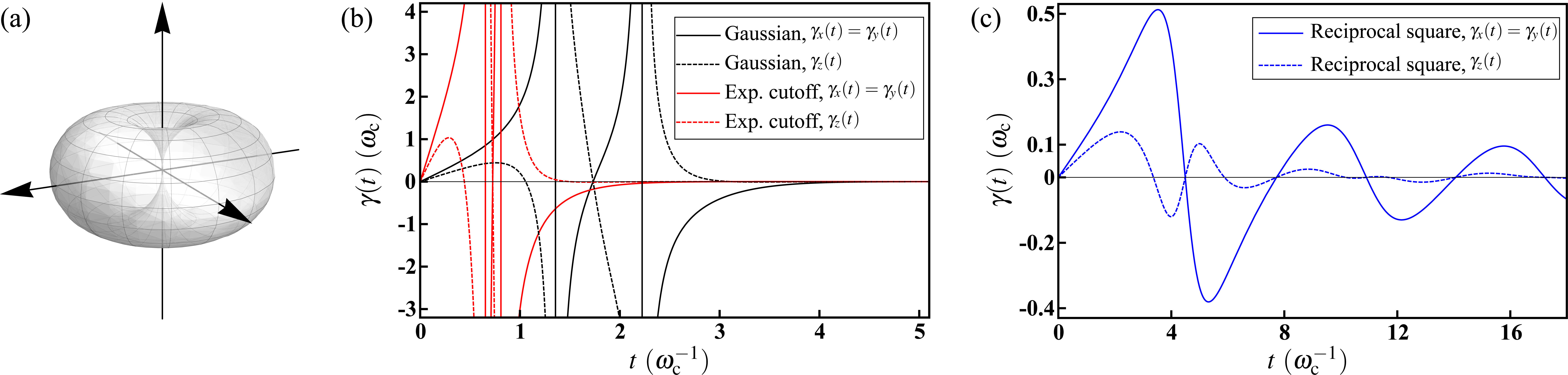}
\caption{Decoherence under the HE with bagel-shaped geometry. (a) Visualization of the solid angular part $\Theta(\theta,\phi)=\pi^{-2}\sin\theta$ of the probability distribution.
The bagel-shaped geometry exhibits the azimuthal and reflectional symmetries simultaneously; meanwhile, this leads to the regime
$\langle n_x^2\rangle_\Theta=\langle n_y^2\rangle_\Theta>1/3>\langle n_z^2\rangle_\Theta$. (b) The decay rates $\gamma_x(t)$ (solid curves) and $\gamma_z(t)$ (dashed curves) for the Gaussian (black)
and the exponential cutoff (red) radial functions. Each $\gamma_x(t)$ exhibits two singularities and finally approaches zero asymptotically. (c) The decay rates $\gamma_x(t)$ (solid curves) and
$\gamma_z(t)$ (dashed curves) for the reciprocal square radial function. $\gamma_x(t)$ shows a similar oscillating behavior; while $\gamma_z(t)$ possesses more zeros. Additionally, the amplitude of
$\gamma_x(t)$ is larger than $\gamma_z(t)$, reflecting the relationship $\langle n_x^2\rangle_\Theta>\langle n_z^2\rangle_\Theta$.}
\label{fig_restults_bagel}
\end{figure*}

We first show the former by considering the case of a bagel-shaped solid angular part with $\Theta(\theta,\phi)=\pi^{-2}\sin\theta$. Its visualization is shown in
Fig.~\ref{fig_restults_bagel}a, from which it is obvious that $\Theta(\theta,\phi)$ exhibits simultaneously the azimuthal symmetry and the reflectional symmetry about the $x$-$y$ plane. With
the specified functional form, we can explicitly compute $\langle n_x^2\rangle_\Theta=\langle n_y^2\rangle_\Theta=3/8$, $\langle n_z^2\rangle_\Theta=1/4$, and $\xi=\int\Theta(\theta,\phi)d\Omega=1$,
satisfying the required relationship $\langle n_x^2\rangle_\Theta=\langle n_y^2\rangle_\Theta>1/3>\langle n_z^2\rangle_\Theta$. In fact, this relationship can also be inferred from the visualization of
$\Theta(\theta,\phi)$ before explicit computations.

\begin{figure*}[ht]
\includegraphics[width=\textwidth]{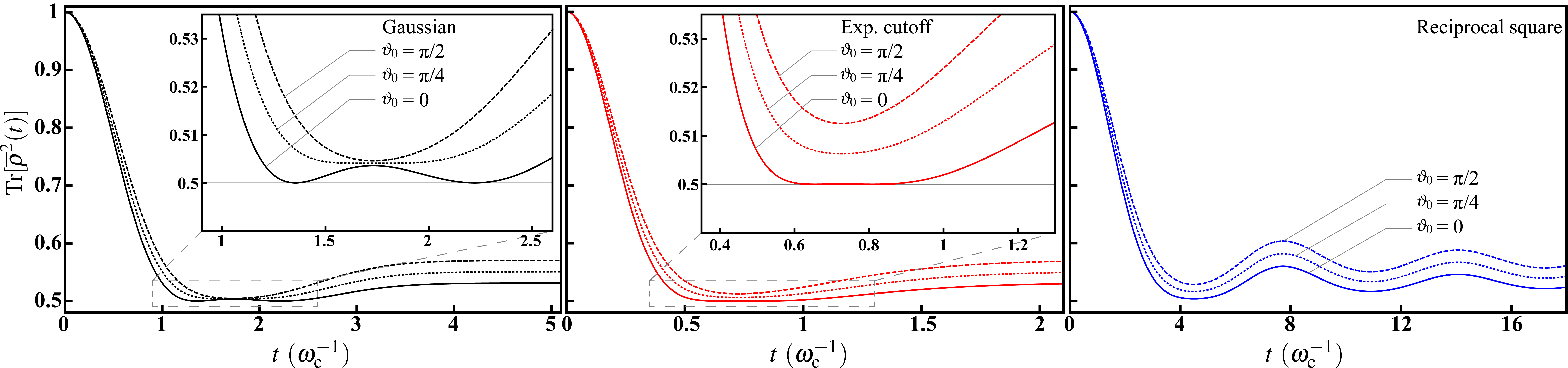}
\caption{The purities $\mathrm{Tr}[\overline{\rho}^2(t)]$ for initial state $\vec{\rho}_0=(\sin\vartheta_0,0,\cos\vartheta_0)$ with $\vartheta_0=0$ (solid curves), $\pi/4$ (dotted curves), and $\pi/2$
(dashed curves), respectively. For $\vartheta_0=0$, strong relation can be observed between the purity and $\gamma_x(t)$ in Fig.~\ref{fig_restults_bagel}. Positive (negative) $\gamma_x(t)$ results in
lowering (rising) purity, respectively; and the singularity of $\gamma_x(t)$ leads to a full die-out of purity followed by a revival. While the case of $\vartheta_0=\pi/2$ is more involved. The profile
of purity is a result of the competition between $\gamma_x(t)$ and $\gamma_z(t)$. Under the regime $\langle n_x^2\rangle_\Theta>\langle n_z^2\rangle_\Theta$, the singular effects of $\gamma_z(t)$ are
quenched by $\gamma_x(t)$ and therefore the purity is always finite.}
\label{fig_restults_purity_bagel}
\end{figure*}

Now we revisit the three radial functions, the Gaussian $P_\mathrm{G}(\omega)$~(\ref{eq_radial_gaussian}), the exponential cutoff $P_\mathrm{EC}(\omega)$~(\ref{eq_radial_exp_cut}), and the reciprocal
square $P_\mathrm{RS}(\omega)$~(\ref{eq_radial_reci_squa}). We first verify that $\int_0^\infty P(\omega)\omega^2 d\omega=\int\Theta(\theta,\phi)d\Omega=1$ guarantees the normalization condition for
$p(\omega,\theta,\phi)$, and the radial expectation $\langle\cos\omega t\rangle_P$ with respect to the three radial functions have been shown in Eqs.~(\ref{eq_expectation_cos_gau}),
(\ref{eq_expectation_cos_exp_cut}), and (\ref{eq_expectation_cos_reci_squa}), respectively. Then the incoherent dynamical behavior can be fully understood and the decay rates $\gamma_j(t)$ in
the master equation~(\ref{eq_aniso_master_eq}) can also be computed explicitly. The analytical expressions of $\gamma_j(t)$ are given in Methods. We show the numerical results with respect
to the radial functions in Fig.~\ref{fig_restults_bagel}.

Figure~\ref{fig_restults_bagel}b shows $\gamma_x(t)$ (solid curves) and $\gamma_z(t)$ (dashed curves) for the Gaussian (black) and the exponential cutoff (red) radial functions. Note that,
each of the sharp descents of $\gamma_x(t)$ for the Gaussian and the exponential cutoff under spherical symmetry [cf. Fig.~(\ref{fig_restults_spherical_symm})b] now splits into two prominent
singularities under lower symmetry. This renders the behavior of $\gamma_z(t)=-\left[\dot{f}_x(t)/f_x(t)\right]-\gamma_x(t)$ singular as well. Finally, they again approach zero asymptotically. The
results for the reciprocal square radial function are shown in Fig.~\ref{fig_restults_bagel}c. In contrast to the former, both $\gamma_x(t)$ and $\gamma_z(t)$ are regular. It can be seen that
$\gamma_x(t)$ shows similar temporal behavior to the one under spherical symmetry [cf. Fig.~(\ref{fig_restults_spherical_symm})b], whereas $\gamma_z(t)$ possesses more zeros. Additionally, it is
interesting to note that the amplitude of $\gamma_x(t)$ is larger than $\gamma_z(t)$, reminiscent of the relationship $\langle n_x^2\rangle_\Theta>\langle n_z^2\rangle_\Theta$ we are considered.
Similar analogy can be observed in the latter example. In fact, this analogy provides some insights into the effects of symmetry breaking on the decay rates, and will be discussed latter.

The numerical results of purity are shown in Fig.~\ref{fig_restults_purity_bagel} for initial state $\vec{\rho}_0=(\sin\vartheta_0,0,\cos\vartheta_0)$ with $\vartheta_0=0$ (solid curves), $\pi/4$
(dotted curves), and $\pi/2$ (dashed curves), respectively. For $\vartheta_0=0$, the time evolution is solely determined by $f_z(t)$ [cf. Eq.~(\ref{eq_dyn_map_pauli_mat_azi_ref_symm})], which in turn
fully determines $\gamma_x(t)$ and $\gamma_y(t)$. Therefore, the profile of purity completely reflects the behavior of $\gamma_x(t)$ in Fig.~\ref{fig_restults_bagel}. It is obvious that positive
(negative) $\gamma_x(t)$ results in lowering (rising) purity, respectively; meanwhile, the singularity of $\gamma_x(t)$ leads to a full die-out of purity followed by a revival. On the other hand, it is
more involved for $\vartheta_0=\pi/2$. Now $f_x(t)$ dominates the time evolution. This leads to a competition between $\gamma_x(t)$ and $\gamma_z(t)$. The singular effects of $\gamma_z(t)$
are quenched by $\gamma_x(t)$ due to the regime $\langle n_x^2\rangle_\Theta>\langle n_z^2\rangle_\Theta$ and therefore the purity is always finite.

\subsection{Dumbbell-shaped solid angular function}
\begin{figure*}[ht]
\includegraphics[width=\textwidth]{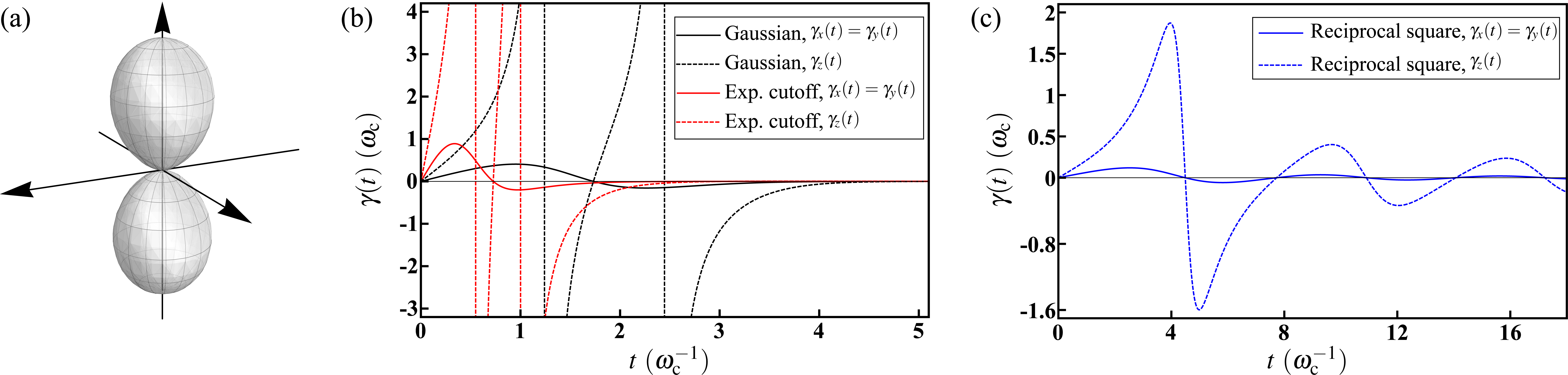}
\caption{Decoherence under the HE with dumbbell-shaped geometry. (a) Visualization of the solid angular part $\Theta(\theta,\phi)=(3/4\pi)\cos^2\theta$ of the probability
distribution. The dumbbell-shaped geometry exhibits the azimuthal and reflectional symmetries simultaneously; meanwhile, this leads to the regime
$\langle n_x^2\rangle_\Theta=\langle n_y^2\rangle_\Theta<1/3<\langle n_z^2\rangle_\Theta$. (b) The decay rates $\gamma_x(t)$ (solid curves) and $\gamma_z(t)$ (dashed curves) for the Gaussian (black)
and the exponential cutoff (red) radial functions. $\gamma_x(t)$'s are finite with smaller amplitudes, whereas $\gamma_z(t)$'s are singular. Furthermore, $\gamma_x(t)$ for the exponential cutoff
approaches zero asymptotically from above after a mild rising. (c) The decay rates $\gamma_x(t)$ (solid curves) and $\gamma_z(t)$ (dashed curves) for the reciprocal square radial function. Both of the
decay rates exhibits the same temporal behavior. Finally, in these plots, we can see the analogy between the amplitudes of $\gamma_j(t)$'s and the relationship
$\langle n_x^2\rangle_\Theta<\langle n_z^2\rangle_\Theta$.}
\label{fig_restults_dumbbell}
\end{figure*}

For the second regime we consider the case of a dumbbell-shaped solid angular part with $\Theta(\theta,\phi)=(3/4\pi)\cos^2\theta$. Its visualization is shown in Fig.~\ref{fig_restults_dumbbell}a
and clearly satisfies the desired simultaneous symmetries. In this case the 2$^\mathrm{nd}$ directional moments are $\langle n_x^2\rangle_\Theta=\langle n_y^2\rangle_\Theta=1/5$,
$\langle n_z^2\rangle_\Theta=3/5$, and $\xi=\int\Theta(\theta,\phi)d\Omega=1$, satisfying the required relationship
$\langle n_x^2\rangle_\Theta=\langle n_y^2\rangle_\Theta<1/3<\langle n_z^2\rangle_\Theta$. Similarly, we adopt the same radial functions again. The analytic expressions are given in
Methods. We show the numerical results in Fig.~\ref{fig_restults_dumbbell}.

\begin{figure*}[ht]
\includegraphics[width=\textwidth]{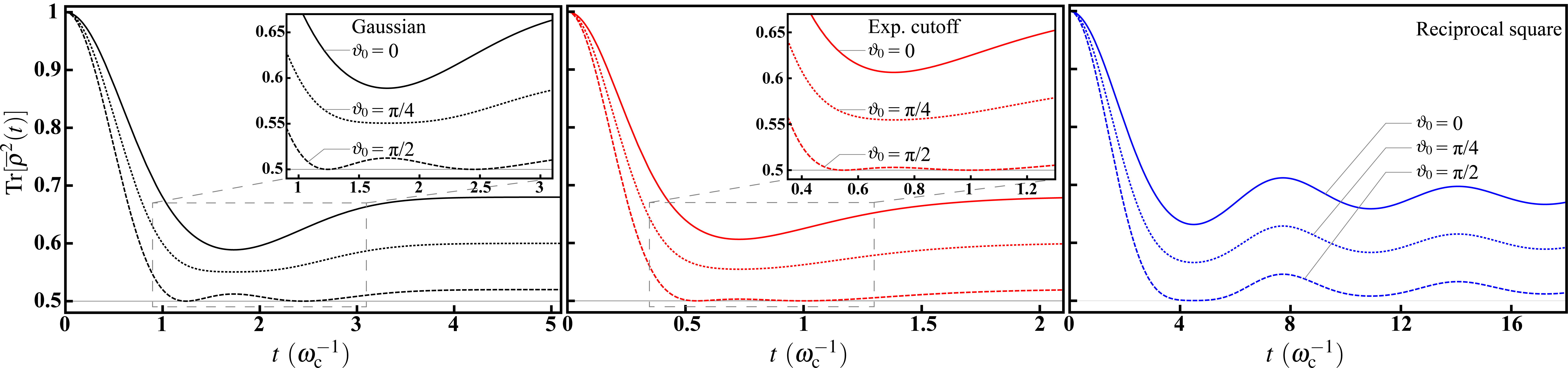}
\caption{The purities $\mathrm{Tr}[\overline{\rho}^2(t)]$ for initial state $\vec{\rho}_0=(\sin\vartheta_0,0,\cos\vartheta_0)$ with $\vartheta_0=0$ (solid curves), $\pi/4$ (dotted curves), and $\pi/2$
(dashed curves), respectively. Aforementioned relation between the purity and $\gamma_j(t)$'s in Fig.~\ref{fig_restults_dumbbell} can also be observed. However, due to the regime
$\langle n_x^2\rangle_\Theta<\langle n_z^2\rangle_\Theta$ considered here, smaller amplitude of $\gamma_x(t)$ implies larger purity for $\vartheta_0=0$. While for the case of $\vartheta_0=\pi/2$, the
singular effects of $\gamma_z(t)$ are dominant and therefore lead to a full die-out of purity.}
\label{fig_restults_purity_dumbbell}
\end{figure*}

Figure.~\ref{fig_restults_dumbbell}b shows the results of $\gamma_x(t)$ (solid curves) and $\gamma_z(t)$ (dashed curves) for the Gaussian (black) and the exponential cutoff (red) radial
functions. Under this regime $\gamma_x(t)$'s are finite with amplitudes smaller than the ones under spherical symmetry [cf. Fig.~\ref{fig_restults_spherical_symm}b], whereas $\gamma_z(t)$'s are still
singular, reflecting the relationship $\langle n_x^2\rangle_\Theta<\langle n_z^2\rangle_\Theta$ as well. On the other hand, $\gamma_x(t)$ for the Gaussian exhibits a similar temporal behavior to the one
under spherical symmetry; while $\gamma_x(t)$ for exponential cutoff exhibits an additional mild rising from negative and finally approaches zero  asymptotically from above. The results for the
reciprocal square radial function are shown in Fig.~\ref{fig_restults_dumbbell}c. We can see that both of the decay rates exhibits the same temporal behavior and, consequently, possess the same
zeros. Meanwhile, We again see the same analogy between the amplitudes of $\gamma_j(t)$'s and the relationship $\langle n_x^2\rangle_\Theta<\langle n_z^2\rangle_\Theta$.

The profile of purity shown in Fig.~\ref{fig_restults_purity_dumbbell} basically obeys the same logic as that in Fig.~\ref{fig_restults_purity_bagel}. However, as here we consider the opposite
regime, smaller amplitude of $\gamma_x(t)$ implies larger purity for $\vartheta_0=0$. The singular effects of $\gamma_z(t)$ are now dominant in the competition with $\gamma_x(t)$ and therefore lead to
a full die-out of purity for $\vartheta_0=\pi/2$.

It has been pointed out that, for qubit pure dephsing, the dephasing rate is dominated by the variance of the probability distribution within HE~\cite{Kropf2016effective}. On the other hand, provided
the simultaneous symmetries considered here, the 2$^\mathrm{nd}$ directional moments play the role of variance alone the specified directions. Consequently, the asymmetry goes into play via them,
as reflected by the decay rates, and gives rise to the analogy. This point can be understood from the decay rates $\gamma_j(t)$ in Eq.~(\ref{eq_aniso_master_eq}), along with $f_j(t)$ in
Eq.~(\ref{eq_f_functions}). Furthermore, according to the symmetries under consideration, we have $\langle n_x^2\rangle_\Theta=\langle n_y^2\rangle_\Theta\neq\langle n_z^2\rangle_\Theta$.
This implies these two regimes and demonstrates striking difference in the dynamical behavior. In the following, we will further reduce the symmetry by considering the azimuthal symmetry exclusively
and explore the effects of 1$^\mathrm{st}$ directional moment.

\section{Azimuthal symmetry}

\begin{figure*}[ht]
\includegraphics[width=\textwidth]{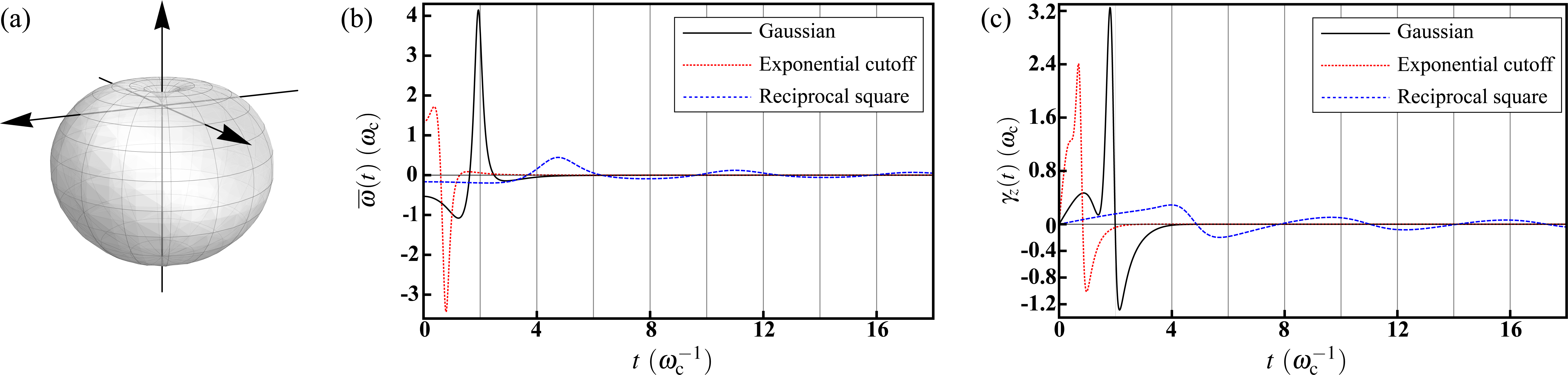}
\caption{Decoherence under the HE with the geometry of 3D cardioid. (a) Visualization of the solid angular part $\Theta(\theta,\phi)=(1-\cos\theta)/4\pi$ of the probability
distribution. This geometry exhibits the azimuthal symmetry exclusively; meanwhile, this leads to the balanced regime
$\langle n_x^2\rangle_\Theta=\langle n_y^2\rangle_\Theta=1/3=\langle n_z^2\rangle_\Theta$. (b) The averaged level spacing $\overline{\omega}(t)$ for the Gaussian (black solid curve), the exponential
cutoff (red dotted curve), and the reciprocal square (blue dashed cruve) radial functions. Its presence is a result of the lack of the reflectional symmetry about the $x$-$y$ plane.
(c) The decay rate $\gamma_z(t)$ for the Gaussian (black solid curve), the exponential cutoff (red dotted curve), and the reciprocal square (blue dashed cruve) radial functions.
The line shape is similar to the one under spherical symmetry appended by an additional drop before the peaking value, which is a result of the asymmetry.}
\label{fig_restults_azi_symm_card}
\end{figure*}

In the presence of the azimuthal symmetry, it has been shown in the previous
examples that the 1$^\mathrm{st}$ $x$- and $y$-directional moments vanish, $\langle n_x\rangle_\Theta=\langle n_y\rangle_\Theta=0$, and the 2$^\mathrm{nd}$ $x$- and $y$-directional moments are equal,
$\langle n_x^2\rangle_\Theta=\langle n_y^2\rangle_\Theta$. Crucially, in the absence of reflectional symmetry about the $x$-$y$ plane, the 1$^\mathrm{st}$ $z$-directional moment is typically
non-vanishing, $\langle n_z\rangle_\Theta\neq0$. For example, we will consider a solid angular part of three-dimensional cardioid with $\Theta(\theta,\phi)=(1-\cos\theta)/4\pi$. Its visualization
is shown in Fig.~\ref{fig_restults_azi_symm_card}a.

According to the directional moments determined by the symmetry under consideration,
the action of ensemble-averaged dynamical linear map in Eqs.~(\ref{eq_dyn_map_pauli_mat}) can be explicitly written as
\begin{equation}
\left\{\begin{array}{l}
\mathcal{E}_t\{\widehat{I}\}=\widehat{I} \\
\mathcal{E}_t\{\hat{\sigma}_x\}=f_x(t)\hat{\sigma}_x+\langle\sin\omega t\rangle_P\langle n_z\rangle_\Theta\hat{\sigma}_y\\
\mathcal{E}_t\{\hat{\sigma}_y\}=f_y(t)\hat{\sigma}_y-\langle\sin\omega t\rangle_P\langle n_z\rangle_\Theta\hat{\sigma}_x\\
\mathcal{E}_t\{\hat{\sigma}_z\}=f_z(t)\hat{\sigma}_z
\end{array}\right.,
\label{eq_dynamical_map_azi_symm}
\end{equation}
In the presence of $\langle n_z\rangle_\Theta$, we have one more additional radial expectation $\langle\sin\omega t\rangle_P$; meanwhile, the incoherent dynamical behavior is governed by the master
equation
\begin{equation}
\frac{\partial}{\partial t}\overline{\rho}(t)=-i\left[\frac{\overline{\omega}(t)}{2}\hat{\sigma}_z,\overline{\rho}(t)\right]
+\sum_{j=1}^3\frac{\gamma_j(t)}{2}\left[\hat{\sigma}_j\overline{\rho}(t)\hat{\sigma}_j-\overline{\rho}(t)\right].
\label{eq_aniso_master_eq_with_level_spac}
\end{equation}
In deriving the above master equation, we have used the facts that $\langle n_x^2\rangle_\Theta=\langle n_y^2\rangle_\Theta$ and $f_x(t)=f_y(t)$; therefore,
$\gamma_x(t)=\gamma_y(t)=-\dot{f}_z(t)/2f_z(t)$. Note that the most prominent difference of Eq.~(\ref{eq_aniso_master_eq_with_level_spac}) from the previous examples
is the presence of an effective level spacing
$\overline{\omega}(t)=\langle n_z\rangle_\Theta\left[f_x(t)(\frac{d}{dt}\langle\sin\omega t\rangle_P)-\dot{f}_x(t)\langle\sin\omega t\rangle_P\right]/D(t)$ with
$D(t)=f^2_x(t)+\langle n_z\rangle_\Theta^2\langle\sin\omega t\rangle_P^2$. Additionally, the decay rate
$\gamma_z(t)=-\left[f_x(t)\dot{f}_x(t)/D(t)\right]-\gamma_x(t)-\left[\langle n_z\rangle_\Theta^2\langle\sin\omega t\rangle_P(\frac{d}{dt}\langle\sin\omega t\rangle_P)/D(t)\right]$
is also altered in the presence of finite $\langle n_z\rangle_\Theta$. Both of $\overline{\omega}(t)$ and the variation of $\gamma_z(t)$ are results of the lack of the reflectional symmetry
about the $x$-$y$ plane.

Instead of the unbalanced regimes considered in the two previous examples, now we demonstrate a balanced one with
$\langle n_x^2\rangle_\Theta=\langle n_y^2\rangle_\Theta=1/3=\langle n_z^2\rangle_\Theta$ to simplify the complexity. We consider a solid angular part of three-dimensional cardioid with
$\Theta(\theta,\phi)=(1-\cos\theta)/4\pi$. Its visualization is shown in Fig.~\ref{fig_restults_azi_symm_card}a. We can verify that the desired balanced regime is satisfied, and compute
$\langle n_x\rangle_\Theta=\langle n_y\rangle_\Theta=0$ and $\langle n_z\rangle_\Theta=-1/3$. With the same radial functions, we can determine the effective level spacing $\overline{\omega}(t)$ and the
decay rate $\gamma_j(t)$ in the master equation~(\ref{eq_aniso_master_eq_with_level_spac}). In addition to $\langle\cos\omega t\rangle_P$, now we need one more radial expectation
$\langle\sin\omega t\rangle_P$. The analytical expresstions with respect to the three radial functions are given in Methods.

Thanks to the balanced regime under consideration, we find that $\gamma_x(t)=\gamma_y(t)$ and they are exactly the same as those under spherical symmetry; while this is not the case for $\gamma_z(t)$
due to the breaking of the reflectional symmetry. We therefore show the numerical results of $\overline{\omega}(t)$ and $\gamma_z(t)$ in Figs.~\ref{fig_restults_azi_symm_card}b and
\ref{fig_restults_azi_symm_card}c, respectively. For the Gaussian radial function (black solid curve), $\overline{\omega}(t)$ begins with a negative value. After a shallow drop, a sharp peak is
followed, then going down to negative again, and finally approaching zero asymptotically from below. For the exponential cutoff (red dotted), the line shape is similar to that of Gaussian but an
overturned one. However, for the reciprocal square (blue dashed curve), $\overline{\omega}(t)$ exhibits a long oscillating tail due to the finite domain of $P_\mathrm{RS}(\omega)$. On the other hand,
it is interesting to note that, the overall line shape of $\gamma_z(t)$ is similar to the one under spherical symmetry, but an additional drop before reaching the peaking value.
The similarity is a consequence of the balanced regime, according to the aforementioned analogy. However, the asymmetry is also influential by perturbing $\gamma_z(t)$.

\section{Simultaneous reflectional symmetries}

\begin{figure*}[ht]
\includegraphics[width=\textwidth]{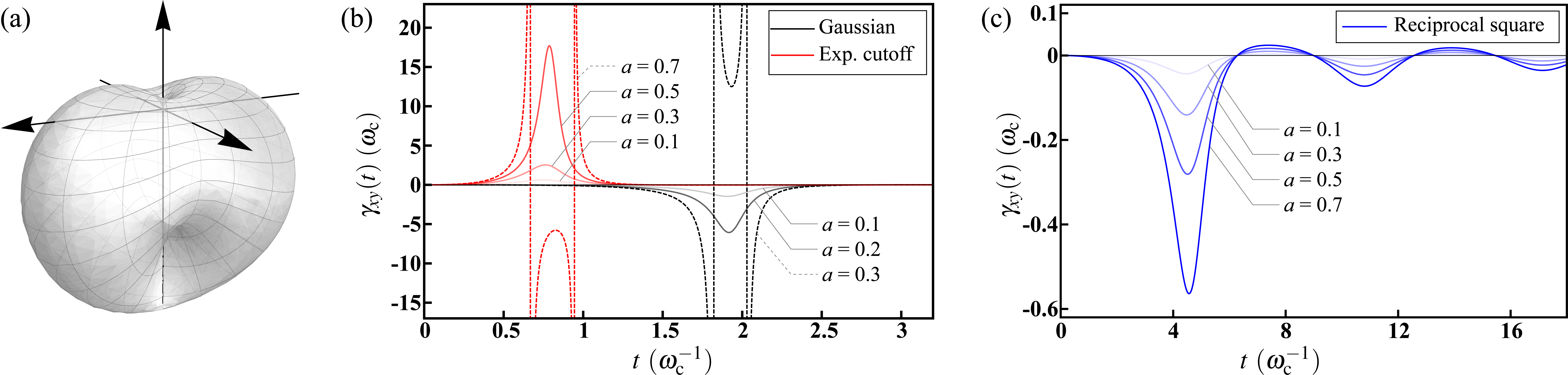}
\caption{Decoherence under the HE with the geometry of kneaded cardioid. (a) Visualization of the solid angular part $\Theta(\theta,\phi)=(1-\cos\theta)(1+a\cos2\phi)/4\pi$ with
lateral asymmetry $a=0.3$. The azimuthal symmetry is broken due to the $\phi$-dependence, and merely the simultaneous reflectional symmetries about the $x$-$z$ and $y$-$z$ planes are left. (b) The decay
rate $\gamma_{xy}(t)$ for the Gaussian (black) and the exponential cutoff (red) radial functions with different lateral asymmetry $a$. As $\gamma_{xy}(t)$ is caused by the azimuthal symmetry breaking,
it is gradually vanishing when $a$ is decreasing and, contrarily, becomes more prominent for a more asymmetric geometry.
As indicated by the dashed curves, the negative peaks even split into two singularities at $a=0.3$ and $0.7$ for Gaussian and exponential cutoff radial functions, respectively. (c) The decay rate
$\gamma_{xy}(t)$ for the reciprocal square radial function with different lateral asymmetry $a$. Similar tendency toward vanishment with decreasing $a$ can also be seen. However, for the reciprocal
square, $\gamma_{xy}(t)$ do not exhibit singularity even if under large asymmetry.}
\label{fig_restults_azi_symm_knead_card}
\end{figure*}

Finally, we consider a more asymmetric case by reducing the continuous azimuthal symmetry to a discrete one, i.e., the simultaneous reflectional symmetries about the $x$-$z$ and $y$-$z$ planes.
In this case, we can only deduce the vanishment of the 1$^\mathrm{st}$ $x$- and $y$-directional moments, $\langle n_x\rangle_\Theta=\langle n_y\rangle_\Theta=0$. The three 2$^\mathrm{nd}$ directional
moments are generically different due to the lack of the azimuthal symmetry.

Given the vanishing 1$^\mathrm{st}$ $x$- and $y$-directional moments, the action of ensemble-averaged dynamical linear map in Eqs.~(\ref{eq_dyn_map_pauli_mat}) is formally the same as Eqs.~(\ref{eq_dynamical_map_azi_symm}) under azimuthal symmetry. Whereas the corresponding master equation
\begin{eqnarray}
\frac{\partial}{\partial t}\overline{\rho}(t)&=&-i\left[\frac{\overline{\omega}(t)}{2}\hat{\sigma}_z,\overline{\rho}(t)\right]
+\sum_{j=1}^3\frac{\gamma_j(t)}{2}\left[\hat{\sigma}_j\overline{\rho}(t)\hat{\sigma}_j-\overline{\rho}(t)\right] \nonumber\\
&&+\frac{\gamma_{xy}(t)}{2}\left[\hat{\sigma}_x\overline{\rho}(t)\hat{\sigma}_y-\frac{1}{2}\left\{\hat{\sigma}_y\hat{\sigma}_x,\overline{\rho}(t)\right\}\right]
+\frac{\gamma_{yx}(t)}{2}\left[\hat{\sigma}_y\overline{\rho}(t)\hat{\sigma}_x-\frac{1}{2}\left\{\hat{\sigma}_x\hat{\sigma}_y,\overline{\rho}(t)\right\}\right]
\label{eq_aniso_master_eq_with_gamma_xy}
\end{eqnarray}
possesses two more decay channels with the off-diagonal decay rates
\begin{equation}
\gamma_{xy}(t)=\gamma_{yx}(t)=-\frac{\langle n_z\rangle_\Theta[f_x(t)-f_y(t)]\frac{d}{dt}\langle\sin\omega t\rangle_P}{2D(t)}
+\frac{\langle n_z\rangle_\Theta[\dot{f}_x(t)-\dot{f}_y(t)]\langle\sin\omega t\rangle_P}{2D(t)},
\label{eq_gamma_xy}
\end{equation}
where $D(t)=f_x(t)f_y(t)+\langle n_z\rangle_\Theta^2\langle\sin\omega t\rangle_P^2$. It is clear that their presences are a result of the azimuthal symmetry breaking. As this is the lowest degree of
symmetry considered here, Eq.~(\ref{eq_aniso_master_eq_with_gamma_xy}) is the most general master equation we have demonstrated. It can be reduced to all the previous ones if the corresponding
symmetries are recovered.

To exemplify this case of low symmetry,  we consider a solid angular part of kneaded cardioid with $\Theta(\theta,\phi)=(1-\cos\theta)(1+a\cos2\phi)/4\pi$, where $0\leq a\leq1$ describes the degree of
lateral asymmetry of the geometry. Its visualization is shown in Fig.~\ref{fig_restults_azi_symm_knead_card}a with $a=0.3$. Due to the $\phi$-dependence, the geometry
appears to be subject to stress along the $y$-axis, and then expanding along the $x$-axis. Therefore the azimuthal symmetry is broken and merely the simultaneous reflectional symmetries are left. For
this $\Theta(\theta,\phi)$, we have $\langle n_z\rangle_\Theta=-1/3$, $\langle n_x^2 \rangle_\Theta=(2+a)/6$, $\langle n_y^2 \rangle_\Theta=(2-a)/6$, and $\langle n_z^2 \rangle_\Theta=1/3$. Along with
the radial expectations $\langle\cos\omega t\rangle_P$ shown in Eqs.~(\ref{eq_expectation_cos_gau}), (\ref{eq_expectation_cos_exp_cut}), and (\ref{eq_expectation_cos_reci_squa}), as well as
$\langle\sin\omega t\rangle_P$ in Methods, the numerical results of $\gamma_{xy}(t)$~(\ref{eq_gamma_xy}) are shown in Fig.~\ref{fig_restults_azi_symm_knead_card}.

The results for the Gaussian radial function are represented in Fig.~\ref{fig_restults_azi_symm_knead_card}b by black (solid and dashed) curves with varying lateral asymmetry $a$. As expected,
$\gamma_{xy}(t)$ is gradually vanishing with decreasing $a$ (i.e., more symmetric geometry), reflecting the fact that $\gamma_{xy}(t)$ is a result of azimuthal symmetry breaking. In contrast,
$\gamma_{xy}(t)$ becomes more prominent when $a$ is large. The negative peak even splits into two singularities at $a=0.3$, as indicated by the black dashed curve. Moreover, the results for the
exponential cutoff radial function are represented by red (solid and dashed) curves with varying $a$. It can be seen that the overall tendency is the same as Gaussian but an overturned one. The curves
show a positive peak at an earlier time; meanwhile, the exponential-cutoff case shows singularities at a larger value of $a=0.7$, as indicated by the red dashed curve.

On the other hand, the results for the reciprocal square radial function are represented in Fig.~\ref{fig_restults_azi_symm_knead_card}c by blue curves with varying $a$. The tendency toward vanishment
with decreasing $a$ can also be seen; whereas, in contrast to the other two radial functions, the deformation of the line shape with $a$ is even, without showing singularity, even if $a$ is large.
Moreover, akin to the other decay rates with the same reciprocal square radial function, $\gamma_{xy}(t)$ also exhibits an oscillating tail, after the negative main peak. Crucially, from these results
of Gaussian and reciprocal square radial functions, there is a distinct property the off-diagonal decay rate $\gamma_{xy}(t)$ can be observed as well. To guarantee the complete positivity of the
dynamical linear map $\mathcal{E}_t$, the short time behavior of all the digonal decay rates $\gamma_j(t)$ associated to the Pauli decay channels is rising at beginning. Any negative values can only
emerge after a period of positive values. However, $\gamma_{xy}(t)$ is possible to have a short time behavior of descent to negative at beginning. We stress that this does not imply the violation of
complete positivity as the eigenvalues of the Kossakowski matrix $\mathcal{K}=[\gamma_{jk}(t)]$ exhibit a short time behavior of rising to positive at beginning.

\section{Conclusion and discussion}

In summary, we have explored the incoherent dynamics of qubits raised from the ensemble average over the canonical HE of structural disorder. Under the variable separation
(\ref{eq_separation_condition}) of the probability distribution within the HE into the radial and the solid angular parts, the structural disorder can be characterized in accordance with the degree of
symmetry of the solid angular geometry. The effects of asymmetry are particularly manifest via the corresponding master equation in the Lindblad form.

In this work we have considered three radial functions; namely, the Gaussian, the exponential cutoff, and the reciprocal square. We first show the most symmetric case of spherical symmetry, leading to
the master equation of isotropic depolarization. It is worthwhile to note that the canonical HE of spherical symmetry not only reproduces the results of unitarily invariant disorder for qubit
\cite{Kropf2016effective}, but also resolves the issue of double counting. Further versatility of the canonical HE in describing various types of incoherent dynamics beyond isotropic depolarization is
revealed by considering asymmetric cases. In addition to the spherical symmetry, we have also demonstrated three lower degrees of symmetry. We can observe the effects caused by the symmetry breaking of
different types via the master equation in the Lindblad form, including the singularities of the decay rates, the effective level spacing, and the off-diagonal decay rates. Generally speaking, the more
asymmetric the geometry is, the more terms emerge.

Notably, the asymmetry goes into play via the 1$^\mathrm{st}$ and 2$^\mathrm{nd}$ directional moments, which in turn determine the aforementioned terms in the master equation. The effective level
spacing $\overline{\omega}(t)$ emerges in the presence of 1$^\mathrm{st}$ directional moments, which are caused by the reflectional symmetry breaking. They denote the mean values alone the specified
directions. While the time dependence of $\overline{\omega}(t)$ is determined by the 2$^\mathrm{nd}$ directional moments and the radial expectations, characterizing the overall asymmetry and the radial
disorder, respectively. On the other hand, the behaviors, particularly the time dependence, of the decay rates are also intimately related to the directional moments. This can be easily understood from
the analogy between the relationship of the 2$^\mathrm{nd}$ directional moments (balanced/unbalanced regimes) and the amplitudes of the decay rates. Moreover, the presence of 1$^\mathrm{st}$ directional
moments will further complicate the decay rates. Finally, the highly asymmetric case gives rise to the emergence of off-diagonal decay rates. Interestingly, very similar conclusions have been drawn
under the framework of qubit pure dephasing \cite{Kropf2016effective}, where it has been shown that $\overline{\omega}(t)$ is given by the mean value and the asymmetry of the distribution, while the
dephasing rate is given by the variance and the kurtosis.

Finally, it is worthwhile to note that the variable separation~(\ref{eq_separation_condition}) can be further released by considering correlated radial and angular coordinates, leading to even more
general joint probability distributions $p(\omega,\theta,\phi)$. For example, expansion in terms of spherical harmonics $\mathrm{Y}_{l,m}(\theta,\phi)$ has been demonstrated in
Ref.~\cite{hongbin_cher_sr_2021}. Furthermore, this problem can also be investigted from the viewpoint of RU, which is highly related to HE.
Since the RU representation is a useful tool in the study of open system dynamics \cite{pernice_randon_unitary_jpb_2012,chruscinski_k_divi_pra_2015,megier_randon_unitary_sr_2017}, considerable efforts
have been devoted into the investigation of UR decomposition \cite{audenaert_randon_unitary_njp_2008,mendl_randon_unitary_cmp_2009,julius_randon_unitary_pra_2009}. For example, master equation in
the same form as Eq.~(\ref{eq_aniso_master_eq}) is derived by mixing fluctuating Gaussian noise along different angles \cite{megier_randon_unitary_sr_2017}.

On the other hand, it is know that any qubit or qutrit phase damping dynamical mpas are RU \cite{landau_randon_unitary_1993,buscemi_err_correction_prl_2005}. However, neither the RU-decomposition nor
the extreme points for higher dimensional cases are unclear. These problems can in general merely be numerically implemented \cite{landau_randon_unitary_1993}. Therefore, the canonical HE would not only
provide insights into the effects of higher asymmetry beyond the framework of existing RU representation, but also establish a systematic approach to single out the attainable subset, and substantially
simplify the problem of RU-decomposition. This underpins the significance of our approach.

($\equiv\widehat{\mathsf{\Phi}}\omega\widehat{\mathsf{\Phi}}\equiv$)$\sim$meow$\sim$ \newline

\section*{Methods}

\subsection*{Orthogonal transform of the 2$^\mathrm{nd}$ directional moments} \label{app_u_transform_2nd_mom}

According to the last term of Eq.~(\ref{eq_unitary_single_realization}) in the main text, there should be 9 terms in the last summation of Eq.~(\ref{eq_averaged_dynamics}):
\begin{equation}
\sum_{j',k'=1}^3\langle n_{j'}n_{k'} \rangle_\Theta\hat{\sigma}_{j'}\rho_0\hat{\sigma}_{k'}.
\end{equation}
Since the 9 terms of 2$^\mathrm{nd}$ directional moment $\langle n_{j'}n_{k'} \rangle_\Theta$ forms a symmetric matrix, it can be diagonalized with an appropriate orthogonal matrix $\hat{u}=[u_{j'j}]$
such that
\begin{equation}
\langle n_{j'}n_{k'} \rangle_\Theta=\sum_{j,k=1}^{3}u_{j'j}\delta_{jk}\langle n_{j}^2\rangle_\Theta u_{k'k}.
\end{equation}
This corresponds to a basis transformation $\hat{\sigma}_j=\sum_{j'=1}^3u_{j'j}\hat{\sigma}_{j'}$ and a rotation $n_j=\sum_{j'=1}^3u_{j'j}n_{j'}$ of the directional unit vector. The later is necessary
in the first summation of Eq.~(\ref{eq_averaged_dynamics}).

\subsection*{Decoherence under simultaneous azimuthal and reflectional symmetries} \label{app_b}

As explained in the main text, given the probability distribution $p(\omega,\theta,\phi)=P(\omega)\Theta(\theta,\phi)$ with the solid angular part $\Theta(\theta,\phi)$ exhibiting simultaneously the
azimuthal symmetry and the reflectional symmetry about the $x$-$y$ plane, we can compute the decay rates
\begin{equation}
\left\{\begin{array}{l}
\gamma_x(t)=\gamma_y(t)=-\dot{f}_z(t)/2f_z(t)\\
\gamma_z(t)=-\left[\dot{f}_x(t)/f_x(t)\right]+\left[\dot{f}_z(t)/2f_z(t)\right]=-\left[\dot{f}_x(t)/f_x(t)\right]-\gamma_x(t)
\end{array}\right.
\end{equation}
in the master equation~(\ref{eq_aniso_master_eq}) of anisotropic depolarization, where
$f_j(t)=\langle\cos\omega t\rangle_P\left(\xi-\langle n_j^2 \rangle_\Theta\right)+\langle n_j^2 \rangle_\Theta/\xi$. In these expressions, we have used the facts that
$\langle n_x^2\rangle_\Theta=\langle n_y^2\rangle_\Theta$ and $f_x(t)=f_y(t)$ due to the azimuthal symmetry.

\subsection*{Decay rates for bagel-shaped geometry} \label{app_c}

Given the solid angular part $\Theta(\theta,\phi)=\pi^{-2}\sin\theta$, we can further compute $\langle n_x^2\rangle_\Theta=\langle n_y^2\rangle_\Theta=3/8$, $\langle n_z^2\rangle_\Theta=1/4$, and
$\xi=\int\Theta(\theta,\phi)d\Omega=1$.

To fully determine the probability distribution $p(\omega,\theta,\phi)=P(\omega)\Theta(\theta,\phi)$, we revisit the three radial functions defined in Eqs.~(\ref{eq_radial_gaussian}), (\ref{eq_radial_exp_cut}), and (\ref{eq_radial_reci_squa}) in the main text. Then the radial expectation $\langle\cos\omega t\rangle_P$ with respect to the three radial
functions are shown in Eqs.~(\ref{eq_expectation_cos_gau}), (\ref{eq_expectation_cos_exp_cut}), and (\ref{eq_expectation_cos_reci_squa}), respectively.
Combining all the necessary information above, we obtain the decay rates for Gaussian $P_\mathrm{G}(\omega)$
\begin{equation}
\left\{\begin{array}{l}
\gamma_x(t)=\gamma_y(t)=3\omega_\mathrm{c}\frac{\omega_\mathrm{c}t\left[3-(\omega_\mathrm{c}t)^2\right]}{6\left[1-(\omega_\mathrm{c}t)^2\right]+2\exp\left[(\omega_\mathrm{c}t)^2/2\right]}\\
\gamma_z(t)=5\omega_\mathrm{c}\frac{\omega_\mathrm{c}t\left[3-(\omega_\mathrm{c}t)^2\right]}{5\left[1-(\omega_\mathrm{c}t)^2\right]+3\exp\left[(\omega_\mathrm{c}t)^2/2\right]}-\gamma_x(t)
\end{array}\right.,
\end{equation}
for exponential cutoff $P_\mathrm{EC}(\omega)$
\begin{equation}
\left\{\begin{array}{l}
\gamma_x(t)=\gamma_y(t)=6\omega_\mathrm{c}^2t\frac{\left[3-(\omega_\mathrm{c}t)^2\right]\left[1+(\omega_\mathrm{c}t)^2\right]+2\left[1-6(\omega_\mathrm{c}t)^2+(\omega_\mathrm{c}t)^4\right]}
{3\left[1-6(\omega_\mathrm{c}t)^2+(\omega_\mathrm{c}t)^4\right]\left[1+(\omega_\mathrm{c}t)^2\right]+\left[1+(\omega_\mathrm{c}t)^2\right]^5}\\
\gamma_z(t)=20\omega_\mathrm{c}^2t\frac{\left[3-(\omega_\mathrm{c}t)^2\right]\left[1+(\omega_\mathrm{c}t)^2\right]+2\left[1-6(\omega_\mathrm{c}t)^2+(\omega_\mathrm{c}t)^4\right]}
{5\left[1-6(\omega_\mathrm{c}t)^2+(\omega_\mathrm{c}t)^4\right]\left[1+(\omega_\mathrm{c}t)^2\right]+3\left[1+(\omega_\mathrm{c}t)^2\right]^5}-\gamma_x(t)
\end{array}\right.,
\end{equation}
and for reciprocal square $P_\mathrm{RS}(\omega)$
\begin{equation}
\left\{\begin{array}{l}
\gamma_x(t)=\gamma_y(t)=3\omega_\mathrm{c}\frac{\sin\omega_\mathrm{c}t-(\omega_\mathrm{c}t)\cos\omega_\mathrm{c}t}{6(\omega_\mathrm{c}t)\sin\omega_\mathrm{c}t+2(\omega_\mathrm{c}t)^2}\\
\gamma_z(t)=5\omega_\mathrm{c}\frac{\sin\omega_\mathrm{c}t-(\omega_\mathrm{c}t)\cos\omega_\mathrm{c}t}{5(\omega_\mathrm{c}t)\sin\omega_\mathrm{c}t+3(\omega_\mathrm{c}t)^2}-\gamma_x(t)
\end{array}\right.,
\end{equation}
respectively. The numerical results are shown in Fig.~\ref{fig_restults_bagel}.

\subsection*{Decay rates for dumbbell-shaped geometry} \label{app_d}

Given the solid angular part $\Theta(\theta,\phi)=(3/4\pi)\cos^2\theta$, we can further compute $\langle n_x^2\rangle_\Theta=\langle n_y^2\rangle_\Theta=1/5$, $\langle n_z^2\rangle_\Theta=3/5$, and
$\xi=\int\Theta(\theta,\phi)d\Omega=1$. Given the same radial functions again, then we have the same radial expectations $\langle\cos\omega t\rangle_P$ as shown in
Eqs.~(\ref{eq_expectation_cos_gau}), (\ref{eq_expectation_cos_exp_cut}), and (\ref{eq_expectation_cos_reci_squa}), respectively. We can compute the decay rates for Gaussian $P_\mathrm{G}(\omega)$
\begin{equation}
\left\{\begin{array}{l}
\gamma_x(t)=\gamma_y(t)=\omega_\mathrm{c}\frac{\omega_\mathrm{c}t\left[3-(\omega_\mathrm{c}t)^2\right]}{2\left[1-(\omega_\mathrm{c}t)^2\right]+3\exp\left[(\omega_\mathrm{c}t)^2/2\right]}\\
\gamma_z(t)=4\omega_\mathrm{c}\frac{\omega_\mathrm{c}t\left[3-(\omega_\mathrm{c}t)^2\right]}{4\left[1-(\omega_\mathrm{c}t)^2\right]+\exp\left[(\omega_\mathrm{c}t)^2/2\right]}-\gamma_x(t)
\end{array}\right.,
\end{equation}
for exponential cutoff $P_\mathrm{EC}(\omega)$
\begin{equation}
\left\{\begin{array}{l}
\gamma_x(t)=\gamma_y(t)=4\omega_\mathrm{c}^2t\frac{\left[3-(\omega_\mathrm{c}t)^2\right]\left[1+(\omega_\mathrm{c}t)^2\right]+2\left[1-6(\omega_\mathrm{c}t)^2+(\omega_\mathrm{c}t)^4\right]}
{2\left[1-6(\omega_\mathrm{c}t)^2+(\omega_\mathrm{c}t)^4\right]\left[1+(\omega_\mathrm{c}t)^2\right]+3\left[1+(\omega_\mathrm{c}t)^2\right]^5}\\
\gamma_z(t)=16\omega_\mathrm{c}^2t\frac{\left[3-(\omega_\mathrm{c}t)^2\right]\left[1+(\omega_\mathrm{c}t)^2\right]+2\left[1-6(\omega_\mathrm{c}t)^2+(\omega_\mathrm{c}t)^4\right]}
{4\left[1-6(\omega_\mathrm{c}t)^2+(\omega_\mathrm{c}t)^4\right]\left[1+(\omega_\mathrm{c}t)^2\right]+\left[1+(\omega_\mathrm{c}t)^2\right]^5}-\gamma_x(t)
\end{array}\right.,
\end{equation}
and for reciprocal square $P_\mathrm{RS}(\omega)$
\begin{equation}
\left\{\begin{array}{l}
\gamma_x(t)=\gamma_y(t)=\omega_\mathrm{c}\frac{\sin\omega_\mathrm{c}t-(\omega_\mathrm{c}t)\cos\omega_\mathrm{c}t}{2(\omega_\mathrm{c}t)\sin\omega_\mathrm{c}t+3(\omega_\mathrm{c}t)^2}\\
\gamma_z(t)=4\omega_\mathrm{c}\frac{\sin\omega_\mathrm{c}t-(\omega_\mathrm{c}t)\cos\omega_\mathrm{c}t}{4(\omega_\mathrm{c}t)\sin\omega_\mathrm{c}t+(\omega_\mathrm{c}t)^2}-\gamma_x(t)
\end{array}\right.,
\end{equation}
respectively. The numerical results are shown in Fig.~\ref{fig_restults_dumbbell}.

\subsection*{Radial expectations for the three radial functions} \label{app_e}

In the absence of the reflectional symmetry, we encounter an additional radial expectation $\langle\sin\omega t\rangle_P$ in Eq.~(\ref{eq_dynamical_map_azi_symm}), which is necessary in
determining the effective level spacing and the decay rates. We can compute $\langle\sin\omega t\rangle_P$ analytically for Gaussian $P_\mathrm{G}(\omega)$

\begin{equation}
\langle\sin\omega t\rangle_{P_\mathrm{G}(\omega)}=\int_0^\infty\sin\omega t P_\mathrm{G}(\omega)\omega^2d\omega=\sqrt{\frac{2}{\pi}}\omega_\mathrm{c}t+
e^{-\frac{(\omega_\mathrm{c}t)^2}{2}}\left[1-(\omega_\mathrm{c}t)^2\right]\mathrm{erfi}\left(\frac{\omega_\mathrm{c}t}{\sqrt{2}}\right),
\end{equation}
for exponential cutoff $P_\mathrm{EC}(\omega)$
\begin{equation}
\langle\sin\omega t\rangle_{P_\mathrm{EC}}=\int_0^\infty\sin\omega t P_\mathrm{EC}(\omega)\omega^2d\omega=\frac{-4\omega_\mathrm{c}t+4(\omega_\mathrm{c}t)^3}{\left[1+(\omega_\mathrm{c}t)^2\right]^4},
\end{equation}
and for reciprocal square $P_\mathrm{RS}(\omega)$
\begin{equation}
\langle\sin\omega t\rangle_{P_\mathrm{RS}}=\int_0^{\omega_\mathrm{c}}\sin\omega t P_\mathrm{RS}(\omega)\omega^2d\omega=\frac{1-\cos\omega_\mathrm{c}t}{\omega_\mathrm{c}t},
\end{equation}
respectively.


\begin{thebibliography}{10}
\urlstyle{rm}
\expandafter\ifx\csname url\endcsname\relax
  \def\url#1{\texttt{#1}}\fi
\expandafter\ifx\csname urlprefix\endcsname\relax\def\urlprefix{URL }\fi
\expandafter\ifx\csname doiprefix\endcsname\relax\def\doiprefix{DOI: }\fi
\providecommand{\bibinfo}[2]{#2}
\providecommand{\eprint}[2][]{\url{#2}}

\bibitem{breuer_textbook}
\bibinfo{author}{Breuer, H.-P.} \& \bibinfo{author}{Petruccione, F.}
\newblock \emph{\bibinfo{title}{The Theory of Open Quantum Systems}}
  (\bibinfo{publisher}{Oxford University Press}, \bibinfo{address}{Oxford},
  \bibinfo{year}{2007}).

\bibitem{weiss_textbook}
\bibinfo{author}{Weiss, U.}
\newblock \emph{\bibinfo{title}{Quantum Dissipative Systems, 4th ed.}}
  (\bibinfo{publisher}{World Scientific}, \bibinfo{address}{Singapore},
  \bibinfo{year}{2012}).

\bibitem{fleming_non_mark_review_annphys_2012}
\bibinfo{author}{Fleming, C.} \& \bibinfo{author}{Hu, B.}
\newblock \bibinfo{journal}{\bibinfo{title}{Non-markovian dynamics of open
  quantum systems: Stochastic equations and their perturbative solutions}}.
\newblock {\emph{\JournalTitle{Ann. Phys.}}} \textbf{\bibinfo{volume}{327}},
  \bibinfo{pages}{1238--1276}, \doiprefix\url{10.1016/j.aop.2011.12.006}
  (\bibinfo{year}{2012}).

\bibitem{ines_non_mark_review_rmp_2017}
\bibinfo{author}{de~Vega, I.} \& \bibinfo{author}{Alonso, D.}
\newblock \bibinfo{journal}{\bibinfo{title}{Dynamics of non-markovian open
  quantum systems}}.
\newblock {\emph{\JournalTitle{Rev. Mod. Phys.}}}
  \textbf{\bibinfo{volume}{89}}, \bibinfo{pages}{015001},
  \doiprefix\url{10.1103/RevModPhys.89.015001} (\bibinfo{year}{2017}).

\bibitem{hongbin_3sbm_scirep_2015}
\bibinfo{author}{Chen, H.-B.}, \bibinfo{author}{Lambert, N.},
  \bibinfo{author}{Cheng, Y.-C.}, \bibinfo{author}{Chen, Y.-N.} \&
  \bibinfo{author}{Nori, F.}
\newblock \bibinfo{journal}{\bibinfo{title}{Using non-markovian measures to
  evaluate quantum master equations for photosynthesis}}.
\newblock {\emph{\JournalTitle{Sci. Rep.}}} \textbf{\bibinfo{volume}{5}},
  \bibinfo{pages}{12753}, \doiprefix\url{10.1038/srep12753}
  (\bibinfo{year}{2015}).

\bibitem{ladd_quant_comp_nature_2010}
\bibinfo{author}{Ladd, T.~D.} \emph{et~al.}
\newblock \bibinfo{journal}{\bibinfo{title}{Quantum computers}}.
\newblock {\emph{\JournalTitle{Nature}}} \textbf{\bibinfo{volume}{464}},
  \bibinfo{pages}{45}, \doiprefix\url{10.1038/nature08812}
  (\bibinfo{year}{2010}).

\bibitem{buluta_quant_comp_rpp_2011}
\bibinfo{author}{Buluta, I.}, \bibinfo{author}{Ashhab, S.} \&
  \bibinfo{author}{Nori, F.}
\newblock \bibinfo{journal}{\bibinfo{title}{Natural and artificial atoms for
  quantum computation}}.
\newblock {\emph{\JournalTitle{Rep. Prog. Phys.}}}
  \textbf{\bibinfo{volume}{74}}, \bibinfo{pages}{104401}
  (\bibinfo{year}{2011}).

\bibitem{preskill_quantum_2018}
\bibinfo{author}{Preskill, J.}
\newblock \bibinfo{journal}{\bibinfo{title}{Quantum {C}omputing in the {NISQ}
  era and beyond}}.
\newblock {\emph{\JournalTitle{{Quantum}}}} \textbf{\bibinfo{volume}{2}},
  \bibinfo{pages}{79}, \doiprefix\url{10.22331/q-2018-08-06-79}
  (\bibinfo{year}{2018}).

\bibitem{barreiro_q_sim_nature_2011}
\bibinfo{author}{Barreiro, J.~T.} \emph{et~al.}
\newblock \bibinfo{journal}{\bibinfo{title}{An open-system quantum simulator
  with trapped ions}}.
\newblock {\emph{\JournalTitle{Nature}}} \textbf{\bibinfo{volume}{470}},
  \bibinfo{pages}{486}, \doiprefix\url{10.1038/nature09801}
  (\bibinfo{year}{2011}).

\bibitem{georgescu_q_sim_rmp_2014}
\bibinfo{author}{Georgescu, I.~M.}, \bibinfo{author}{Ashhab, S.} \&
  \bibinfo{author}{Nori, F.}
\newblock \bibinfo{journal}{\bibinfo{title}{Quantum simulation}}.
\newblock {\emph{\JournalTitle{Rev. Mod. Phys.}}}
  \textbf{\bibinfo{volume}{86}}, \bibinfo{pages}{153--185},
  \doiprefix\url{10.1103/RevModPhys.86.153} (\bibinfo{year}{2014}).

\bibitem{keesling_q_sim_nature_2019}
\bibinfo{author}{Keesling, A.} \emph{et~al.}
\newblock \bibinfo{journal}{\bibinfo{title}{Quantum Kibble–Zurek mechanism
  and critical dynamics on a programmable Rydberg simulator}}.
\newblock {\emph{\JournalTitle{Nature}}} \textbf{\bibinfo{volume}{568}},
  \bibinfo{pages}{207}, \doiprefix\url{10.1038/s41586-019-1070-1}
  (\bibinfo{year}{2019}).

\bibitem{scully_qhe_pnas_2011}
\bibinfo{author}{Scully, M.~O.}, \bibinfo{author}{Chapin, K.~R.},
  \bibinfo{author}{Dorfman, K.~E.}, \bibinfo{author}{Kim, M.~B.} \&
  \bibinfo{author}{Svidzinsky, A.}
\newblock \bibinfo{journal}{\bibinfo{title}{Quantum heat engine power can be
  increased by noise-induced coherence}}.
\newblock {\emph{\JournalTitle{Proc. Natl. Acad. Sci. U.S.A.}}}
  \textbf{\bibinfo{volume}{108}}, \bibinfo{pages}{15097--15100},
  \doiprefix\url{10.1073/pnas.1110234108} (\bibinfo{year}{2011}).

\bibitem{rossnage_1_atom_qhe_science_2016}
\bibinfo{author}{Ro{\ss}nagel, J.} \emph{et~al.}
\newblock \bibinfo{journal}{\bibinfo{title}{A single-atom heat engine}}.
\newblock {\emph{\JournalTitle{Science}}} \textbf{\bibinfo{volume}{352}},
  \bibinfo{pages}{325--329}, \doiprefix\url{10.1126/science.aad6320}
  (\bibinfo{year}{2016}).

\bibitem{hongbin_vib_coh_bio_qhe_2016}
\bibinfo{author}{Chen, H.-B.}, \bibinfo{author}{Chiu, P.-Y.} \&
  \bibinfo{author}{Chen, Y.-N.}
\newblock \bibinfo{journal}{\bibinfo{title}{Vibration-induced coherence
  enhancement of the performance of a biological quantum heat engine}}.
\newblock {\emph{\JournalTitle{Phys. Rev. E}}} \textbf{\bibinfo{volume}{94}},
  \bibinfo{pages}{052101}, \doiprefix\url{10.1103/PhysRevE.94.052101}
  (\bibinfo{year}{2016}).

\bibitem{kossakowski_cptp_osid_2003}
\bibinfo{author}{Kossakowski, A.}
\newblock \bibinfo{journal}{\bibinfo{title}{A class of linear positive maps in
  matrix algebras}}.
\newblock {\emph{\JournalTitle{Open Syst. Info. Dyn.}}}
  \textbf{\bibinfo{volume}{10}}, \bibinfo{pages}{213--220},
  \doiprefix\url{10.1023/A:1025101606680} (\bibinfo{year}{2003}).

\bibitem{benatti_open_system_dyna_ijmpb_2005}
\bibinfo{author}{Benatti, F.} \& \bibinfo{author}{Floreanini, R.}
\newblock \bibinfo{journal}{\bibinfo{title}{Open quantum dynamics: Complete
  positivity and entanglement}}.
\newblock {\emph{\JournalTitle{Int. J. Mod. Phys. B}}}
  \textbf{\bibinfo{volume}{19}}, \bibinfo{pages}{3063--3139},
  \doiprefix\url{10.1142/S0217979205032097} (\bibinfo{year}{2005}).

\bibitem{dominy_cptp_maps_qip_2016}
\bibinfo{author}{Dominy, J.~M.}, \bibinfo{author}{Shabani, A.} \&
  \bibinfo{author}{Lidar, D.~A.}
\newblock \bibinfo{journal}{\bibinfo{title}{A general framework for complete
  positivity}}.
\newblock {\emph{\JournalTitle{Quant. Info. Proc.}}}
  \textbf{\bibinfo{volume}{15}}, \bibinfo{pages}{465--494},
  \doiprefix\url{10.1007/s11128-015-1148-0} (\bibinfo{year}{2016}).

\bibitem{chruscinski_open_system_dyna_osid_2017}
\bibinfo{author}{Chru\ifmmode \acute{s}\else
  \'{s}\fi{}ci\ifmmode~\acute{n}\else \'{n}\fi{}ski, D.} \&
  \bibinfo{author}{Pascazio, S.}
\newblock \bibinfo{journal}{\bibinfo{title}{A brief history of the GLKS
  equation}}.
\newblock {\emph{\JournalTitle{Open Syst. Info. Dyn.}}}
  \textbf{\bibinfo{volume}{24}}, \bibinfo{pages}{1740001},
  \doiprefix\url{10.1142/S1230161217400017} (\bibinfo{year}{2017}).

\bibitem{kraus_textbook}
\bibinfo{author}{Kraus, K.}
\newblock \emph{\bibinfo{title}{States, Effects, and Operations: Fundamental
  Notions of Quantum Theory}} (\bibinfo{publisher}{Springer},
  \bibinfo{address}{Berlin, Heidelberg}, \bibinfo{year}{1983}).

\bibitem{man_qi_text_book}
\bibinfo{author}{Nielsen, M.~A.} \& \bibinfo{author}{Chuang, I.~L.}
\newblock \emph{\bibinfo{title}{Quantum Compution and Quantum Information}}
  (\bibinfo{publisher}{Cambridge University Press},
  \bibinfo{address}{Cambridge, England}, \bibinfo{year}{2000}).

\bibitem{jamiolkowski_rmp_1972}
\bibinfo{author}{Jamio{\l}kowski, A.}
\newblock \bibinfo{journal}{\bibinfo{title}{Linear transformations which
  preserve trace and positive semidefiniteness of operators}}.
\newblock {\emph{\JournalTitle{Rep. Math. Phys.}}}
  \textbf{\bibinfo{volume}{3}}, \bibinfo{pages}{275 -- 278},
  \doiprefix\url{10.1016/0034-4877(72)90011-0} (\bibinfo{year}{1972}).

\bibitem{choi_laa_1975}
\bibinfo{author}{Choi, M.-D.}
\newblock \bibinfo{journal}{\bibinfo{title}{Completely positive linear maps on
  complex matrices}}.
\newblock {\emph{\JournalTitle{Linear Alg. Appl.}}}
  \textbf{\bibinfo{volume}{10}}, \bibinfo{pages}{285 -- 290},
  \doiprefix\url{10.1016/0024-3795(75)90075-0} (\bibinfo{year}{1975}).

\bibitem{Kropf2016effective}
\bibinfo{author}{Kropf, C.~M.}, \bibinfo{author}{Gneiting, C.} \&
  \bibinfo{author}{Buchleitner, A.}
\newblock \bibinfo{journal}{\bibinfo{title}{Effective dynamics of disordered
  quantum systems}}.
\newblock {\emph{\JournalTitle{Phys. Rev. X}}} \textbf{\bibinfo{volume}{6}},
  \bibinfo{pages}{031023}, \doiprefix\url{10.1103/PhysRevX.6.031023}
  (\bibinfo{year}{2016}).

\bibitem{hongbin_process_n_cla_prl_2018}
\bibinfo{author}{Chen, H.-B.}, \bibinfo{author}{Gneiting, C.},
  \bibinfo{author}{Lo, P.-Y.}, \bibinfo{author}{Chen, Y.-N.} \&
  \bibinfo{author}{Nori, F.}
\newblock \bibinfo{journal}{\bibinfo{title}{Simulating open quantum systems
  with hamiltonian ensembles and the nonclassicality of the dynamics}}.
\newblock {\emph{\JournalTitle{Phys. Rev. Lett.}}}
  \textbf{\bibinfo{volume}{120}}, \bibinfo{pages}{030403},
  \doiprefix\url{10.1103/PhysRevLett.120.030403} (\bibinfo{year}{2018}).

\bibitem{Gneiting2016incoherent}
\bibinfo{author}{Gneiting, C.}, \bibinfo{author}{Anger, F.~R.} \&
  \bibinfo{author}{Buchleitner, A.}
\newblock \bibinfo{journal}{\bibinfo{title}{Incoherent ensemble dynamics in
  disordered systems}}.
\newblock {\emph{\JournalTitle{Phys. Rev. A}}} \textbf{\bibinfo{volume}{93}},
  \bibinfo{pages}{032139}, \doiprefix\url{10.1103/PhysRevA.93.032139}
  (\bibinfo{year}{2016}).

\bibitem{kropf2017effective}
\bibinfo{author}{Kropf, C.~M.}, \bibinfo{author}{Shatokhin, V.~N.} \&
  \bibinfo{author}{Buchleitner, A.}
\newblock \bibinfo{journal}{\bibinfo{title}{Open system model for quantum
  dynamical maps with classical noise and corresponding master equations}}.
\newblock {\emph{\JournalTitle{Open Syst. Inf. Dyn.}}}
  \textbf{\bibinfo{volume}{24}}, \bibinfo{pages}{1740012},
  \doiprefix\url{10.1142/S1230161217400121} (\bibinfo{year}{2017}).

\bibitem{gneiting_disordered_pra_2017}
\bibinfo{author}{Gneiting, C.} \& \bibinfo{author}{Nori, F.}
\newblock \bibinfo{journal}{\bibinfo{title}{Quantum evolution in disordered
  transport}}.
\newblock {\emph{\JournalTitle{Phys. Rev. A}}} \textbf{\bibinfo{volume}{96}},
  \bibinfo{pages}{022135}, \doiprefix\url{10.1103/PhysRevA.96.022135}
  (\bibinfo{year}{2017}).

\bibitem{kropf_disordered_prr_2020}
\bibinfo{author}{Kropf, C.~M.}
\newblock \bibinfo{journal}{\bibinfo{title}{Protecting quantum coherences from
  static noise and disorder}}.
\newblock {\emph{\JournalTitle{Phys. Rev. Research}}}
  \textbf{\bibinfo{volume}{2}}, \bibinfo{pages}{033311},
  \doiprefix\url{10.1103/PhysRevResearch.2.033311} (\bibinfo{year}{2020}).

\bibitem{hongbin_process_n_cla_nc_2019}
\bibinfo{author}{Chen, H.-B.} \emph{et~al.}
\newblock \bibinfo{journal}{\bibinfo{title}{Quantifying the nonclassicality of
  pure dephasing}}.
\newblock {\emph{\JournalTitle{Nat. Commun.}}} \textbf{\bibinfo{volume}{10}},
  \bibinfo{pages}{3794}, \doiprefix\url{10.1038/s41467-019-11502-4}
  (\bibinfo{year}{2019}).

\bibitem{hongbin_cher_sr_2021}
\bibinfo{author}{Chen, H.-B.} \& \bibinfo{author}{Chen, Y.-N.}
\newblock \bibinfo{journal}{\bibinfo{title}{Canonical hamiltonian ensemble
  representation of dephasing dynamics and the impact of thermal fluctuations
  on quantum-to-classical transition}}.
\newblock {\emph{\JournalTitle{Sci. Rep.}}} \textbf{\bibinfo{volume}{11}},
  \bibinfo{pages}{10046}, \doiprefix\url{10.1038/s41598-021-89400-3}
  (\bibinfo{year}{2021}).

\bibitem{pernice_randon_unitary_jpb_2012}
\bibinfo{author}{Pernice, A.}, \bibinfo{author}{Helm, J.} \&
  \bibinfo{author}{Strunz, W.~T.}
\newblock \bibinfo{journal}{\bibinfo{title}{System{\textendash}environment
  correlations and non-markovian dynamics}}.
\newblock {\emph{\JournalTitle{J. Phys. B: At. Mol. Opt. Phys.}}}
  \textbf{\bibinfo{volume}{45}}, \bibinfo{pages}{154005},
  \doiprefix\url{10.1088/0953-4075/45/15/154005} (\bibinfo{year}{2012}).

\bibitem{lindblad_jmp_1976}
\bibinfo{author}{Gorini, V.}, \bibinfo{author}{Kossakowski, A.} \&
  \bibinfo{author}{Sudarshan, E. C.~G.}
\newblock \bibinfo{journal}{\bibinfo{title}{Completely positive dynamical
  semigroups of $N$-level systems}}.
\newblock {\emph{\JournalTitle{J. Math. Phys.}}} \textbf{\bibinfo{volume}{17}},
  \bibinfo{pages}{821--825}, \doiprefix\url{10.1063/1.522979}
  (\bibinfo{year}{1976}).

\bibitem{andersson_kraus_decompo_jmo_2007}
\bibinfo{author}{Andersson, E.}, \bibinfo{author}{Cresser, J.~D.} \&
  \bibinfo{author}{Hall, M. J.~W.}
\newblock \bibinfo{journal}{\bibinfo{title}{Finding the kraus decomposition
  from a master equation and vice versa}}.
\newblock {\emph{\JournalTitle{J. Mod. Opt.}}} \textbf{\bibinfo{volume}{54}},
  \bibinfo{pages}{1695--1716}, \doiprefix\url{10.1080/09500340701352581}
  (\bibinfo{year}{2007}).

\bibitem{hongbin_k_divi_pra_2015}
\bibinfo{author}{Chen, H.-B.}, \bibinfo{author}{Lien, J.-Y.},
  \bibinfo{author}{Chen, G.-Y.} \& \bibinfo{author}{Chen, Y.-N.}
\newblock \bibinfo{journal}{\bibinfo{title}{Hierarchy of non-markovianity and
  $k$-divisibility phase diagram of quantum processes in open systems}}.
\newblock {\emph{\JournalTitle{Phys. Rev. A}}} \textbf{\bibinfo{volume}{92}},
  \bibinfo{pages}{042105}, \doiprefix\url{10.1103/PhysRevA.92.042105}
  (\bibinfo{year}{2015}).

\bibitem{chruscinski_k_divi_prl_2014}
\bibinfo{author}{Chru\ifmmode \acute{s}\else
  \'{s}\fi{}ci\ifmmode~\acute{n}\else \'{n}\fi{}ski, D.} \&
  \bibinfo{author}{Maniscalco, S.}
\newblock \bibinfo{journal}{\bibinfo{title}{Degree of non-markovianity of
  quantum evolution}}.
\newblock {\emph{\JournalTitle{Phys. Rev. Lett.}}}
  \textbf{\bibinfo{volume}{112}}, \bibinfo{pages}{120404},
  \doiprefix\url{10.1103/PhysRevLett.112.120404} (\bibinfo{year}{2014}).

\bibitem{bae_k_divi_prl_2016}
\bibinfo{author}{Bae, J.} \& \bibinfo{author}{Chru\ifmmode \acute{s}\else
  \'{s}\fi{}ci\ifmmode~\acute{n}\else \'{n}\fi{}ski, D.}
\newblock \bibinfo{journal}{\bibinfo{title}{Operational characterization of
  divisibility of dynamical maps}}.
\newblock {\emph{\JournalTitle{Phys. Rev. Lett.}}}
  \textbf{\bibinfo{volume}{117}}, \bibinfo{pages}{050403},
  \doiprefix\url{10.1103/PhysRevLett.117.050403} (\bibinfo{year}{2016}).

\bibitem{hongbin_n_mark_pra_2017}
\bibinfo{author}{Chen, H.-B.}, \bibinfo{author}{Chen, G.-Y.} \&
  \bibinfo{author}{Chen, Y.-N.}
\newblock \bibinfo{journal}{\bibinfo{title}{Thermodynamic description of
  non-markovian information flux of nonequilibrium open quantum systems}}.
\newblock {\emph{\JournalTitle{Phys. Rev. A}}} \textbf{\bibinfo{volume}{96}},
  \bibinfo{pages}{062114}, \doiprefix\url{10.1103/PhysRevA.96.062114}
  (\bibinfo{year}{2017}).

\bibitem{chruscinski_k_divi_pra_2015}
\bibinfo{author}{Chru\ifmmode \acute{s}\else
  \'{s}\fi{}ci\ifmmode~\acute{n}\else \'{n}\fi{}ski, D.} \&
  \bibinfo{author}{Wudarski, F.~A.}
\newblock \bibinfo{journal}{\bibinfo{title}{Non-markovianity degree for random
  unitary evolution}}.
\newblock {\emph{\JournalTitle{Phys. Rev. A}}} \textbf{\bibinfo{volume}{91}},
  \bibinfo{pages}{012104}, \doiprefix\url{10.1103/PhysRevA.91.012104}
  (\bibinfo{year}{2015}).

\bibitem{megier_randon_unitary_sr_2017}
\bibinfo{author}{Megier, N.}, \bibinfo{author}{Chru\ifmmode \acute{s}\else
  \'{s}\fi{}ci\ifmmode~\acute{n}\else \'{n}\fi{}ski, D.},
  \bibinfo{author}{Piilo, J.} \& \bibinfo{author}{Strunz, W.~T.}
\newblock \bibinfo{journal}{\bibinfo{title}{Eternal non-markovianity: from
  random unitary to markov chain realisations}}.
\newblock {\emph{\JournalTitle{Sci. Rep.}}} \textbf{\bibinfo{volume}{7}},
  \bibinfo{pages}{6379}, \doiprefix\url{10.1038/s41598-017-06059-5}
  (\bibinfo{year}{2017}).

\bibitem{audenaert_randon_unitary_njp_2008}
\bibinfo{author}{Audenaert, K. M.~R.} \& \bibinfo{author}{Scheel, S.}
\newblock \bibinfo{journal}{\bibinfo{title}{On random unitary channels}}.
\newblock {\emph{\JournalTitle{New J. Phys.}}} \textbf{\bibinfo{volume}{10}},
  \bibinfo{pages}{023011}, \doiprefix\url{10.1088/1367-2630/10/2/023011}
  (\bibinfo{year}{2008}).

\bibitem{mendl_randon_unitary_cmp_2009}
\bibinfo{author}{Mendl, C.~B.} \& \bibinfo{author}{Wolf, M.~M.}
\newblock \bibinfo{journal}{\bibinfo{title}{Unital quantum channels – convex
  structure and revivals of Birkhoff’s theorem}}.
\newblock {\emph{\JournalTitle{Commun. Math. Phys.}}}
  \textbf{\bibinfo{volume}{289}}, \bibinfo{pages}{1057},
  \doiprefix\url{10.1007/s00220-009-0824-2} (\bibinfo{year}{2009}).

\bibitem{julius_randon_unitary_pra_2009}
\bibinfo{author}{Helm, J.} \& \bibinfo{author}{Strunz, W.~T.}
\newblock \bibinfo{journal}{\bibinfo{title}{Quantum decoherence of two
  qubits}}.
\newblock {\emph{\JournalTitle{Phys. Rev. A}}} \textbf{\bibinfo{volume}{80}},
  \bibinfo{pages}{042108}, \doiprefix\url{10.1103/PhysRevA.80.042108}
  (\bibinfo{year}{2009}).

\bibitem{landau_randon_unitary_1993}
\bibinfo{author}{Landau, L.~J.} \& \bibinfo{author}{Streater, R.~F.}
\newblock \bibinfo{journal}{\bibinfo{title}{On Birkhoff's theorem for doubly
  stochastic completely positive maps of matrix algebras}}.
\newblock {\emph{\JournalTitle{Linear Alg. Appl.}}}
  \textbf{\bibinfo{volume}{193}}, \bibinfo{pages}{107--127},
  \doiprefix\url{doi.org/10.1016/0024-3795(93)90274-R} (\bibinfo{year}{1993}).

\bibitem{buscemi_err_correction_prl_2005}
\bibinfo{author}{Buscemi, F.}, \bibinfo{author}{Chiribella, G.} \&
  \bibinfo{author}{Mauro~D'Ariano, G.}
\newblock \bibinfo{journal}{\bibinfo{title}{Inverting quantum decoherence by
  classical feedback from the environment}}.
\newblock {\emph{\JournalTitle{Phys. Rev. Lett.}}}
  \textbf{\bibinfo{volume}{95}}, \bibinfo{pages}{090501},
  \doiprefix\url{10.1103/PhysRevLett.95.090501} (\bibinfo{year}{2005}).

\end{thebibliography}

\section*{Acknowledgments}

This work is supported by the Ministry of Science and Technology, Taiwan, Grants No. MOST 109-2112-M-006-012 and MOST 110-2112-M-006-012,
and partially by Higher Education Sprout Project, Ministry of Education to the Headquarters of University Advancement at NCKU.

\section*{Author contributions}

H.-B.C. is the sole author responsible for all the content of this work.

\section*{Competing interests}

The author declares no competing interests.

\end{document}